# Combined Multi-Plane Phase Retrieval and Super-Resolution Optical Fluctuation Imaging for 4D Cell Microscopy

A. Descloux[1#], K. S. Grußmayer[1#], E. Bostan[2,3], T. Lukes[1], A. Bouwens[1], A. Sharipov[1], S. Geissbuehler[1], A.-L. Mahul-Mellier[4], H. A. Lashuel[4], M. Leutenegger[1,5] and T. Lasser[1*]

[1]École Polytechnique Fédérale de Lausanne, Laboratoire d'Optique Biomédicale, 1015 Lausanne, Switzerland
[2]École Polytechnique Fédérale de Lausanne, Biomedical Imaging Group, 1015 Lausanne, Switzerland
[4] École Polytechnique Fédérale de Lausanne, Laboratory of Molecular and Chemical Biology of Neurodegeneration, Brain Mind Institute, 1015 Lausanne, Switzerland

E. Bostan: Current Address
[3]University of California, Berkeley, Computational Imaging Lab, Berkeley, CA 94720, U. S. A.
M. Leutenegger: Current Address
[5]Max-Planck Institute for Biophysical Chemistry, Department of NanoBiophotonics, 37077 Göttingen, Germany

[#]These authors contributed equally to this work.
[*]T. Lasser email: theo.lasser@epfl.ch

## Abstract

Super-resolution fluorescence microscopy provides unprecedented insight into cellular and subcellular structures. However, going `beyond the diffraction barrier' comes at a price since most far-field super-resolution imaging techniques trade temporal for spatial super-resolution. We propose the combination of a novel label-free white light quantitative phase tomography with fluorescence imaging to provide high-speed imaging and spatial super-resolution. The non-iterative phase reconstruction relies on the acquisition of single images at each z-location and thus enables straightforward 3D phase imaging using a classical microscope. We realized multi-plane imaging using a customized prism for the simultaneous acquisition of 8 planes. This allowed us to not only image live cells in 3D at up to 200 Hz, but also to integrate fluorescence super-resolution optical fluctuation imaging within the same optical instrument. This 4D microscope platform unifies the sensitivity and high temporal resolution of phase tomography with the specificity and high spatial resolution of fluorescence imaging.

**Introduction**

During the last two decades novel wide-field fluorescence imaging techniques ((d)STORM, PALM, etc.)[1–6] based on the localization of a stochastically activated subset of single molecules overcame the diffraction limit and provided new insights into subcellular structures. These techniques require long image sequences containing at least several hundred raw images. Therefore, the gain in spatial resolution comes at the cost of reduced time-resolution[7]. Super-resolution optical fluctuation imaging (SOFI)[8,9] provides an elegant alternative for super-resolution imaging using higher-order cumulant statistics. SOFI is well suited for 3D live cell imaging with a moderate time resolution of up to ~1 second per reconstructed image in a multi-plane microscope[10]. In addition, SOFI tolerates high labeling densities[11,12] and can provide a quantitative assessment of molecular parameters[13,14].

Many biological processes, e.g. membrane blebbing, neuronal signaling or cardiovascular cell dynamics, take place on a sub-second to sub-millisecond timescale[15,16].  As investigation of cell physiology demands long-term, fast imaging potentially unattainable for single molecule localization microscopy[17], a novel imaging concept is needed. Our Phase Retrieval Instrument with Super-resolution Microscopy (PRISM) satisfies these needs as a versatile multi-plane platform with high image acquisition rates, integrating 3D fluorescence imaging and a novel white-light phase tomography. It enables multimodal 4D imaging, combining the molecular specificity and spatial resolution of fluorescence super-resolution with the sensitivity and high-speed of quantitative phase imaging.

Phase imaging is a label-free approach, enabling long-term time-lapse imaging of cellular dynamics.  In 1969, E. Wolf proposed a solution how to infer the refractive index distribution, i.e. cell morphology, from scattered light acquisitions[18]. This seminal work launched a new field known as quantitative phase imaging (QPI), which aims at determining the optical phase delay induced by the refractive index distribution. The phase differences can be assessed with various techniques[19] such as measurement of the interference of the scattered field with a reference field (off-axis holography, HPM)[20–23], controlled phase-shift of the reference field (FPM, SLIM)[24,25] or by measuring defocused image planes (TIE)[26–29]. Several QPI concepts have been extended towards 3D live cell imaging[30–32], where the phase tomogram is acquired by z-scanning the sample or through successive projection measurements with multiple illumination angles[33].

In the present work, we retrieve the 3D quantitative phase from a stack of brightfield images using a simple Fourier filtering. We explain this white-light phase tomography based on the theory of 3D partially coherent image formation. The broad versatility of our technique is demonstrated by a number of cell images. We first extract the high-resolution 3D phase information of fixed cells from a large stack of z-displaced intensities. We then present PRISM, a multi-plane microscope for the simultaneous acquisition of 8 planes. This novel configuration was used for monitoring cell dynamics at up to 200 Hz. Finally, we sequentially image cell samples with 3D-SOFI and phase tomography.

PRISM is conceived as an add-on to existing wide-field microscopes, enabling straightforward implementation of fast quantitative phase tomography and 3D fluorescence super-resolution imaging, overall presenting a unique opportunity to study the complex spatial and temporal physiology of live cells.

**Results**

***3D white light partially coherent image formation***

E. Wolf's solution[18] of the inverse light scattering problem is based on the Helmholtz wave equation and a Green's function ansatz which we used as the underlying theoretical framework for our tomographic 3D phase retrieval. The interaction between a plane wave $U_i(\boldsymbol{x}; \omega, \boldsymbol{k_i})$[1] and a weakly scattering object induces a scattered field $U_s(\boldsymbol{x}; \omega, \boldsymbol{k_i}, \boldsymbol{k})$ which encodes the object's scattering potential $F(\boldsymbol{x})$ (Fig. 1a). The illumination field is decomposed into plane wave components and fully characterized by its angular spectrum. Assuming a single plane wave and supposing weak elastic light scattering (using the first order Born approximation[34]), we obtain an adequate model for the light-object interaction (see Supplementary Material Section 1.1). Taking into account the polychromatic illumination with its spectrum $S(\omega)$, we embedded our analysis in the framework of the generalized Wiener-Khintchine theorem[35]. Accordingly, we acquire an intensity encoding the interference of the scattered field with the illumination field (Fig. 1a). This results in a linear relation

$$\Gamma(\boldsymbol{g}) = \mathrm{i}F(\boldsymbol{g})H(\boldsymbol{g}) \tag{1}$$

showing that the cross-spectral density[35] $\Gamma(\boldsymbol{g})$ is given as the product of the scattering potential $F(\boldsymbol{g})$ and the 3D partially coherent transfer function $H(\boldsymbol{g})$ in Fourier space (for the full derivation including underlying approximations, see Supplementary Material Section 1.2).

---

[1] Cartesian space $\boldsymbol{x} = (x, y, z)$, Fourier space $\boldsymbol{g} = (g_\perp, g_z)$, field frequency $\boldsymbol{\omega}$, illumination wavevector $\boldsymbol{k_i}$, scattered wavevector $\boldsymbol{k}$

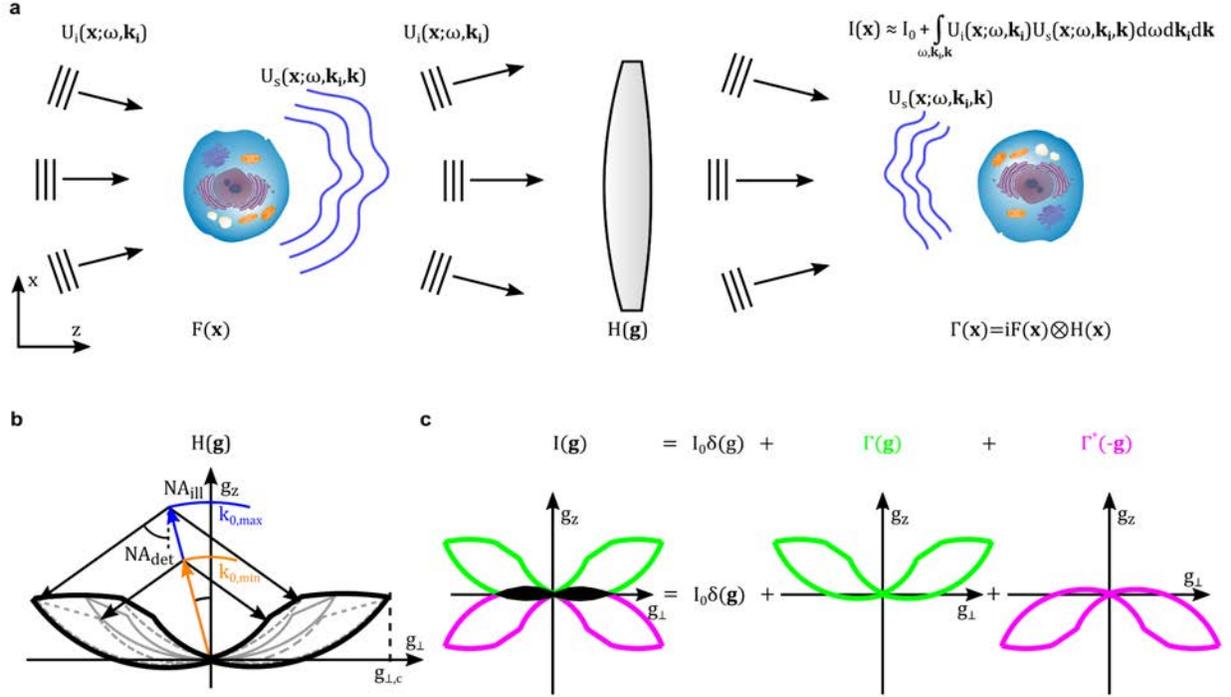

Figure 1: Image formation in partial coherent imaging. (a) An incident plane wave $U_i$ interacts with a weakly scattering object i.e. its scattering potential F. The scattered $U_s$ and unscattered fields are filtered by the optical system H and their interference is acquired at the image plane, (b) Polychromatic transfer function H for Koehler illumination based on an Ewald sphere construction. NA numerical aperture, (c) Fourier representation of the intensity of the interfering fields at the image plane; Γ cross-spectral density; the spatial frequency overlap is indicated in black. Note: The Ewald sphere is a geometric construct for visualizing elastic light scattering. Therefore, the incident and all scattered wavevectors have the same length.

Based on the Ewald sphere construction, this complex interaction is illustrated in Fig. 1b, allowing an intuitive interpretation of elastic light scattering. For each source wavelength $\lambda$, the frequency support corresponds to a shifted sphere cap of radius $k_0 = \bar{n}\frac{2\pi}{\lambda}$, where $\bar{n}$ is the mean refractive index of the sample. The lateral spatial frequencies extend until the cut-off frequency $g_{\perp,c} = k_0(NA_{ill} + NA_{det})$ and depend on the numerical aperture of the Koehler illumination ($NA_{ill}$) and the detection objective ($NA_{det}$) . The final transfer function takes into account the full temporal and angular illumination spectrum (see Supplementary Material Section 1.2, Fig. S1).

Finally, we separate the 3D intensity $I(\boldsymbol{g})$ as the sum of a DC-component and an AC-component, where the AC-component contains the cross-spectral density $\Gamma(\boldsymbol{g})$ and its complex conjugate (Fig. 1c, Supplementary Material Section 2)

$$I(\boldsymbol{g}) \approx I_0\delta(\boldsymbol{g}) + \Gamma(\boldsymbol{g}) + \Gamma^*(-\boldsymbol{g}).$$  (2)

***3D phase tomography***

The experimental procedure is straightforward and consists of acquiring N z-displaced intensities $I(x, y, z_p)$. The brightfield image stack encodes the 3D scattering potential as the interference between the scattered field and the incident field. Referring to the Ewald sphere (Fig. 1b), the attainable frequency support is restricted by the illumination and detection aperture and the illumination spectrum.

Using a low illumination $NA_{ill}$ ($NA_{ill} \leq 0.26$), the cross-spectral density is mainly contained in the upper half Fourier space (see Fig. 1b,c)[36,37]. As indicated in Fig.2a, the cutoff filter $K(g_z) = \begin{cases} 1 & g_z > g_{z,c} \\ 0 & else \end{cases}$ selects $\Gamma_+(\boldsymbol{g}) = I(\boldsymbol{g})K(g_z)$ from the Fourier transformed intensity stack, where the cutoff frequency $g_{z,c}$ is determined by the illumination aperture and the source spectrum (see Supplementary Material Section 2). This filtering implies several far-reaching consequences. First, the cutoff filter allows selecting the complex cross-spectral density $\Gamma_+(\boldsymbol{g})$. Second, $K(g_z)$ suppresses the spectral overlap occurring at low axial frequencies. A simple estimation of the consequences caused by this filtering implies that cell imaging (typical cell dimensions $< 15\,\mu m$) should not be affected by this suppression of low spatial frequencies. Finally, filtering of the spatial frequency support results in optical sectioning, i.e. axial high-pass filtering of the 3D image spectrum. After an inverse 3D Fourier transformation, the quantitative phase associated to the image field is retrieved by normalizing the imaginary part of the cross-spectral density as (Fig. S2, see Supplementary Material section 2)

$$\varphi(\boldsymbol{x}) = \tan^{-1}\left(\frac{Im(\Gamma_+(\boldsymbol{x}))}{I_0 + Re(\Gamma_+(\boldsymbol{x}))}\right). \tag{3}$$

The 3D phase recovery is fast and non-iterative as it only involves linear filtering operations. The method requires just N+1 intensity images to retrieve the 3D phase information of N planes. The common-path interferometric imaging integrating white-light illumination and high detection $NA_{det}$ ($NA_{det} = 1.2$) provides high phase sensitivity and high spatial resolution. We validated the phase retrieval by imaging sparsely distributed polystyrene beads embedded in agarose gel. Our experimental data match well the simulated phase and amplitude (see Supplementary Materials Section 3, Fig. S3). These bead measurements are complemented by assessing the topology of a technical sample (using only 8 through-focus brightfield planes, see Supplementary Materials Section 4, Fig. S4). An independent Atomic Force Microscopy (AFM) measurement confirms the quantitative nature of our phase recovery algorithm.

Fig. 2a summarizes the workflow of our method. The simple algorithm for our forward model results in 3D real time imaging (average processing time of the algorithm optimized for phase accuracy: 0.39 s for a stack of 480 x 480 x 8 pixels on an Intel® Xeon® E5-2650 v2 @ 2.60 GHz).

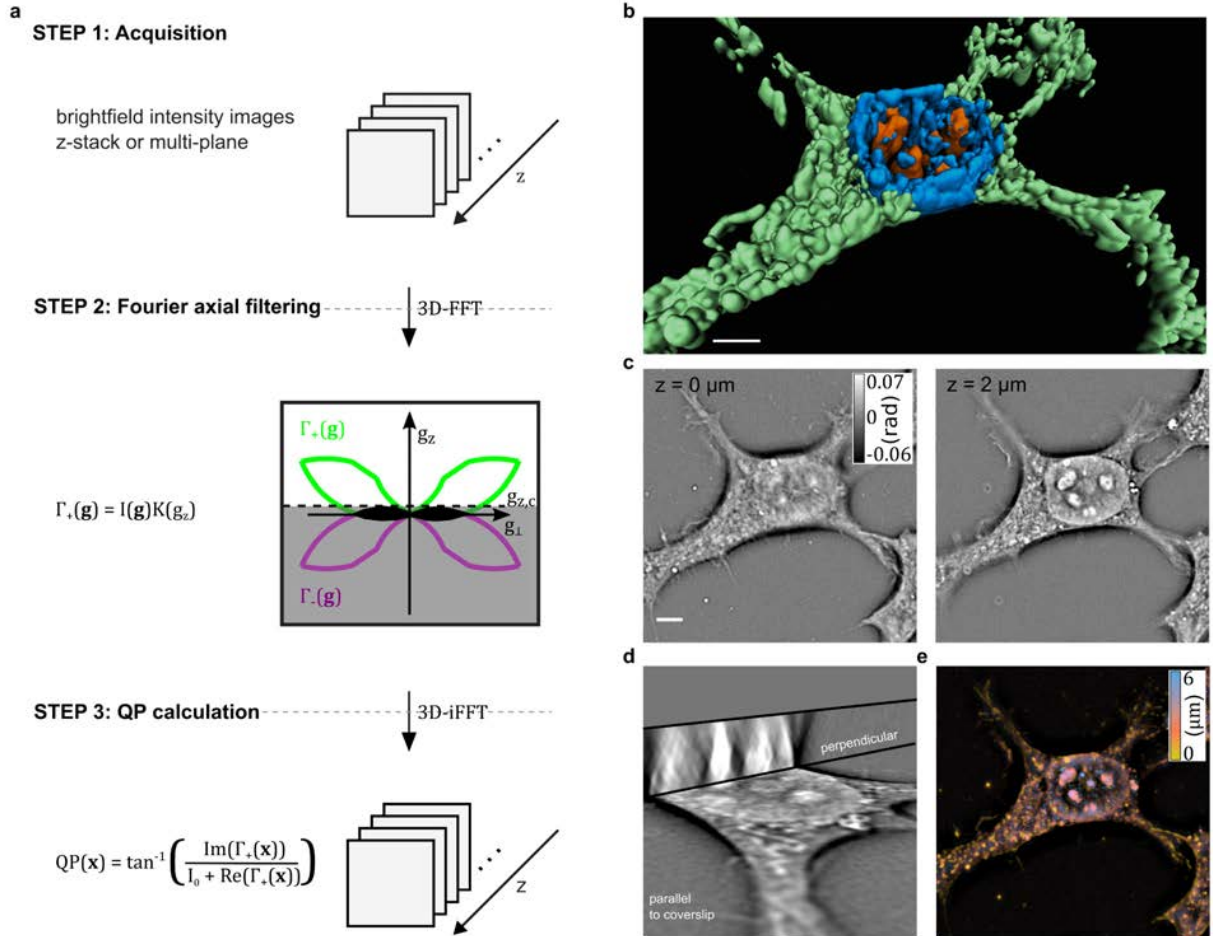

Figure 2: White-light cross-spectral density algorithm for 3D phase imaging. (a) Algorithm workflow comprising bright-field image stack acquisition, Fourier axial filtering and 3D phase retrieval, (b) 3D quantitative phase imaging using 50 planes acquired with a 20 ms camera exposure time and with an inter-plane spacing of 200 nm. Rendered, false-colored tomogram (Imaris, Bitplane) of a fixed HeLa cell along with (c) selected xy-slices, (d) orthogonal views and (e) a color-coded maximum z-projection (ImageJ[38], phase threshold T = 0). Details see Supplementary materials. Scalebar 5 µm.

We first demonstrate cellular imaging on fixed, well-known HeLa human epitheloid cervix carcinoma cells, as summarized in Fig. 2. The high-resolution 3D phase information was recovered from a z-stack of 50 bright-field images acquired with a 20 ms camera exposure time and with a displacement step of 200 nm. In Fig. 2b, the 3D rendered, false-color coded tomogram of an adherent cell is shown. The refractive index variations together with the known morphology of subcellular structures allow segmentation of the cellular outline and the cell nucleus containing the high refractive index nucleoli (cell contour in green; nucleus in blue; nucleoli in orange). Selected xy-slices underline the optical sectioning capability and reveal additional intracellular detail (Fig. 2c). Most likely, we observe organelles such as mitochondria and vesicles; even thin filopodia are clearly visible on the glass substrate. Orthogonal views and a color-coded maximum z-projection illustrate 3D imaging over a depth of 6 µm (Fig. 2d, e). Additional phase tomograms of a variety of different cells types reveals strikingly different morphologies (HEK 293T cells, (stimulated) murine macrophages, mouse hippocampal primary neurons and human fibroblast, see Supplementary Material Fig. S18).

### *PRISM multi-plane platform*

In order to perform fast 3D image acquisition, we conceived a multi-plane platform (MP) based on a novel image splitter. As shown in Fig. 3, a customized image splitting prism in the detection arm of the microscope directs the light into 8 distinct images (see Supplementary Materials Section 6, Table S1 and Fig. S6-S9). The image splitter, placed in the convergent beam path, provides high inter-plane image stability for recording a sample volume of about 50 µm x 50 µm x 2.5 µm. The MP platform allows diffraction limited multiplexed image acquisition of 8 planes with no moving parts (optical design provided in Supplementary Materials Section 6.2 and 6.3, Fig. S10-S16). The dominant speed limitation is given by the camera frame-rate (up to 200 Hz in this work).

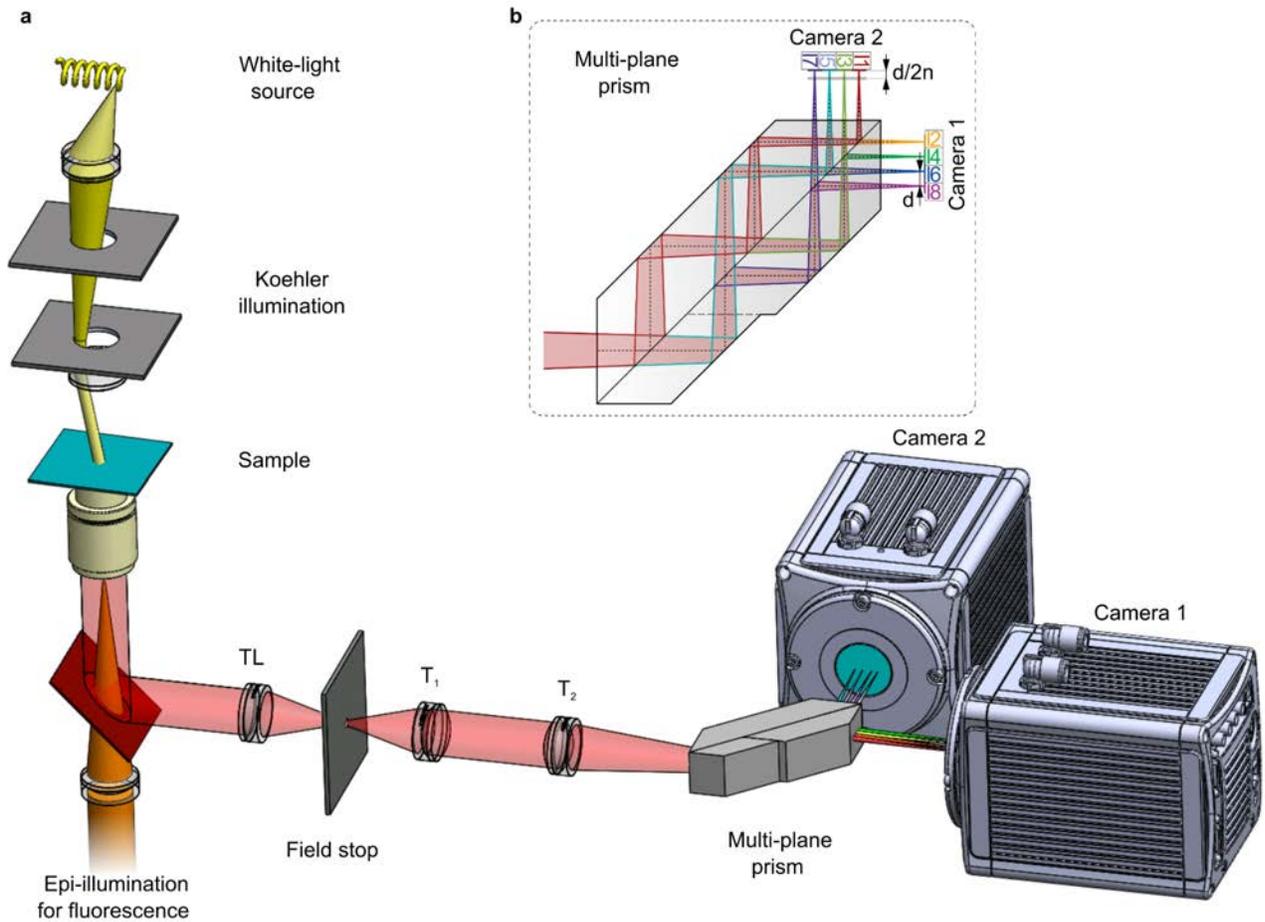

Figure 3: Overview of the PRISM setup. (a) Microscope layout combining epi-fluorescence illumination and white-light Koehler illumination for phase tomography with the multi-plane imaging platform. TL tube lens, T1 and T2 telescope lenses. (b) Multi-plane image splitter (for all specifications and parameters see Supplementary Materials Section 6).

Due to the common path configuration and fast 3D (8 plane) image acquisition, excellent phase stability is achieved. Our high-NA water immersion objective provides a lateral resolution of 380 nm and an axial resolution of 560 nm for phase imaging ($NA_{det} = 1.2$, see Supplementary Materials, Section 3, Fig. S3). We assessed the influence of the 3D intensity sampling on the phase retrieval in detail. As the prism allows fast imaging of 8 planes with an inter-plane distance of 350 nm, we have a trade-off on $g_z$-sampling in Fourier space (see Supplementary Materials section 5, Fig. S5). We have chosen this compromise for the benefit of 3D phase tomography at 200 Hz.

Due to the classical layout of the microscope, adding an epi-illumination fluorescence channel is straightforward (see Fig. 3). The combination of both imaging modalities merges the advantages of phase microscopy such as label-free, long-term dynamic cell imaging with the complementary features of fluorescence, e.g. molecular specificity and single molecule detection. We named our multi-plane microscope Phase Retrieval Instrument with Super-resolution Microscopy (PRISM), as wide-field super-resolution imaging can be easily integrated via the fluorescence channel.

PRISM is ideally suited for 3D Super-resolution Optical Fluctuation Imaging (SOFI). Fast, background-free, super-resolution imaging with inherent optical sectioning is possible by analyzing time-series of independent, stochastically blinking emitters with higher-order cumulant statistics[8]. SOFI can be applied on the same dataset as (d)STORM and PALM[11,39], and also provides reliable results for high labeling densities and limited photon budget. Cross-correlation (or precisely cross-cumulant) analysis is not limited to the lateral dimension. In consequence, if the inter-plane distance is adjusted smaller than the axial PSF extent and the 3D image stack is acquired simultaneously, a 3D super-resolved image acquisition can be realized (see Supplementary Material Section 7 and Fig. S17). Simultaneous multi-plane acquisition significantly reduces the overall imaging time and thus the photobleaching compared to sequential recording of image stacks. We demonstrated a lateral resolution of 110 nm and an axial resolution better than 500 nm for 3rd order 3D SOFI[10]. The chosen specifications for the new MP configuration allows a robust multiplexed image acquisition matching all these requirements. Using PRISM, we demonstrate high-speed live cell phase tomography and the combination of SOFI with phase imaging in 3D.

### *High-speed dynamic 3D phase imaging*

To highlight the fast acquisition of PRISM 3D phase imaging, we monitored a living human fibroblast at an imaging speed of 200 Hz as it migrates on a glass substrate (see Supplementary Movie M1). The overview in Fig. 4a shows the cell body with the nucleus and lamellipodia extending into the direction of migration for a selected plane. The membrane of the nuclear envelope separating the nucleus from the cytoplasm is clearly visible. In the zoomed region of interest (green-dashed square in Fig. 4a), the fast movement of a vesicle (white circle) and an apparent fusion of two small organelles (white arrow) are indicated (Fig. 4b). A kymograph perpendicular to the leading edge of the cell (along the magenta line in Fig. 4a) shows membrane ruffles that move centripetally towards the main body[41] (Fig. 4c). Intracellular vesicle movement in 3D can be observed in the close-up of a color-coded maximum phase z-projection of the green-dashed region of interest (Fig 4d, the particle that is indicated moves up and down (yellow-red-yellow)). We also performed long term 3D imaging of a dividing HeLa cell (Supplementary Material Fig. S19, Supplementary Movie M2) undergoing mitosis from metaphase to telophase, highlighting the stability of the phase imaging over an extended period. These experiments demonstrate the performance of our fast and stable label-free phase tomography.

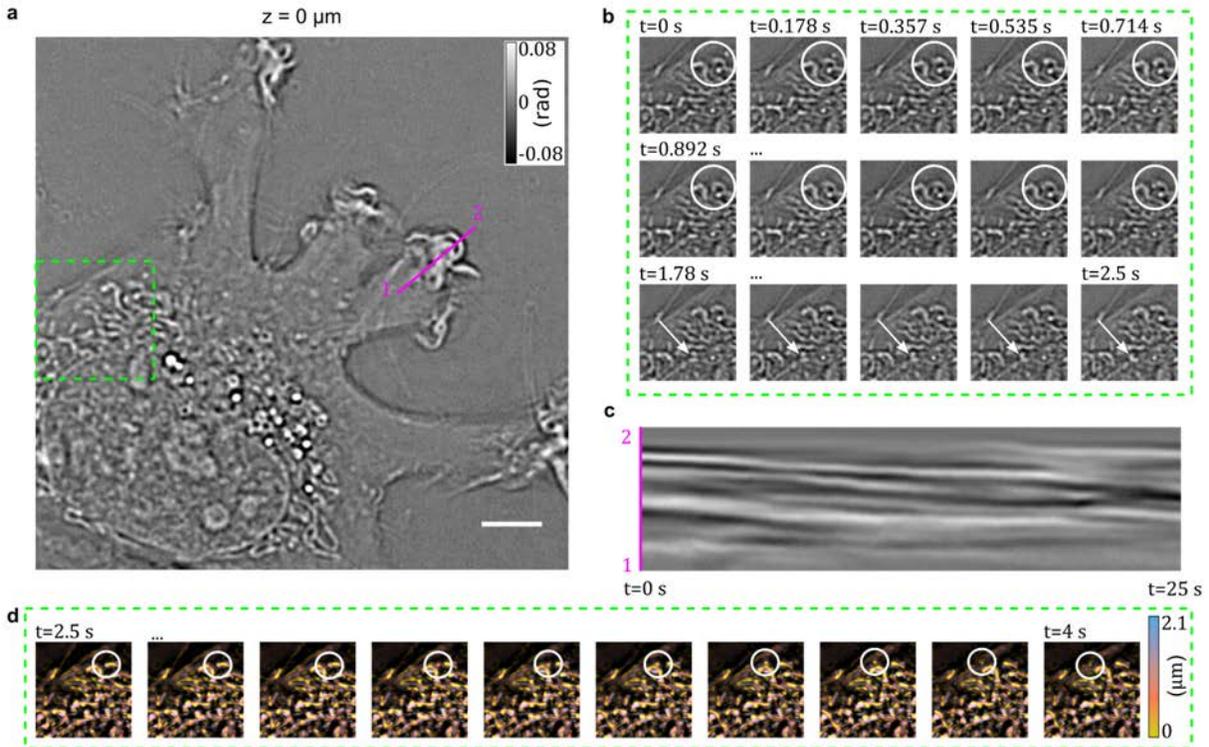

Figure 4: Fast live cell 3D phase imaging of cellular dynamics. (a) Human fibroblast migrating on a glass substrate. First frame of a 25s movie imaged at 200 Hz for a selected plane. (b) Close-up of the green-dashed ROI indicated in (a) at different times illustrates intracellular organelle movement, indicated by a white circle and a white arrow. (c) Kymograph for every 10th frame along the magenta line 1-2 in (a) shows ruffling of lamellipodia. (d) Close-up of the green-dashed ROI indicated in (a) across all planes at different times illustrates intracellular organelle movement in axial direction, indicated by a white circle. Phase color-coded maximum z-projection (threshold T = 0 rad). Scalebar 5 μm.

### Combining SOFI and phase imaging

We explored the full potential of our PRISM multimodal multi-plane imaging by investigating different cell types using super-resolution optical fluctuation imaging followed by phase imaging, both in 3D (Fig. 5).

As a proof of principle, we first acquired image sequences of microtubules in fixed HeLa cells that were fluorescently labeled with Alexa Fluor 647 by indirect immunostaining. We obtained 3D super-resolved images by computing second and third order bSOFI[13] images of the fluctuating signals of blinking labels (Fig. 5a). The optical sectioning capability and the removal of out-of-focus background are apparent in the bSOFI images. A selected xy-plane displays the rich (intra-) cellular context provided by the subsequently acquired corresponding phase tomogram (Fig. 5d). Clearly, the cell nucleus coincides with the void volume that is surrounded by the labeled cytoskeleton filaments, whose network extends almost to the outline of the cell. A color-coded maximum z-projection of the green-dashed region of interest indicated in the 2D phase illustrates the 3D nature of the phase images (Fig. 5g). Our correlative phase and bSOFI images are intrinsically co-aligned, as they were taken successively using the MP microscope without moving the sample.

Fig. 5b, e and h summarize the imaging of mouse hippocampal primary neurons that were treated with α-synuclein fibrils. The protein is abundant in the brain and abnormal accumulation of aggregates is a characteristic for a number of neurodegenerative diseases including Parkinson's[42]. SOFI reveals the 3D architecture of newly formed Alexa Fluor 647-immunostained α-synuclein aggregates (Fig. 5b). Several long fibers extend over the whole imaging depth. The corresponding phase tomogram (Fig. 5e, h) shows that, in this case, most of the α-synuclein aggregates are found within a bundle of neurites. The outline of a neuronal cell body is barely visible slightly off the center of the 2D image.

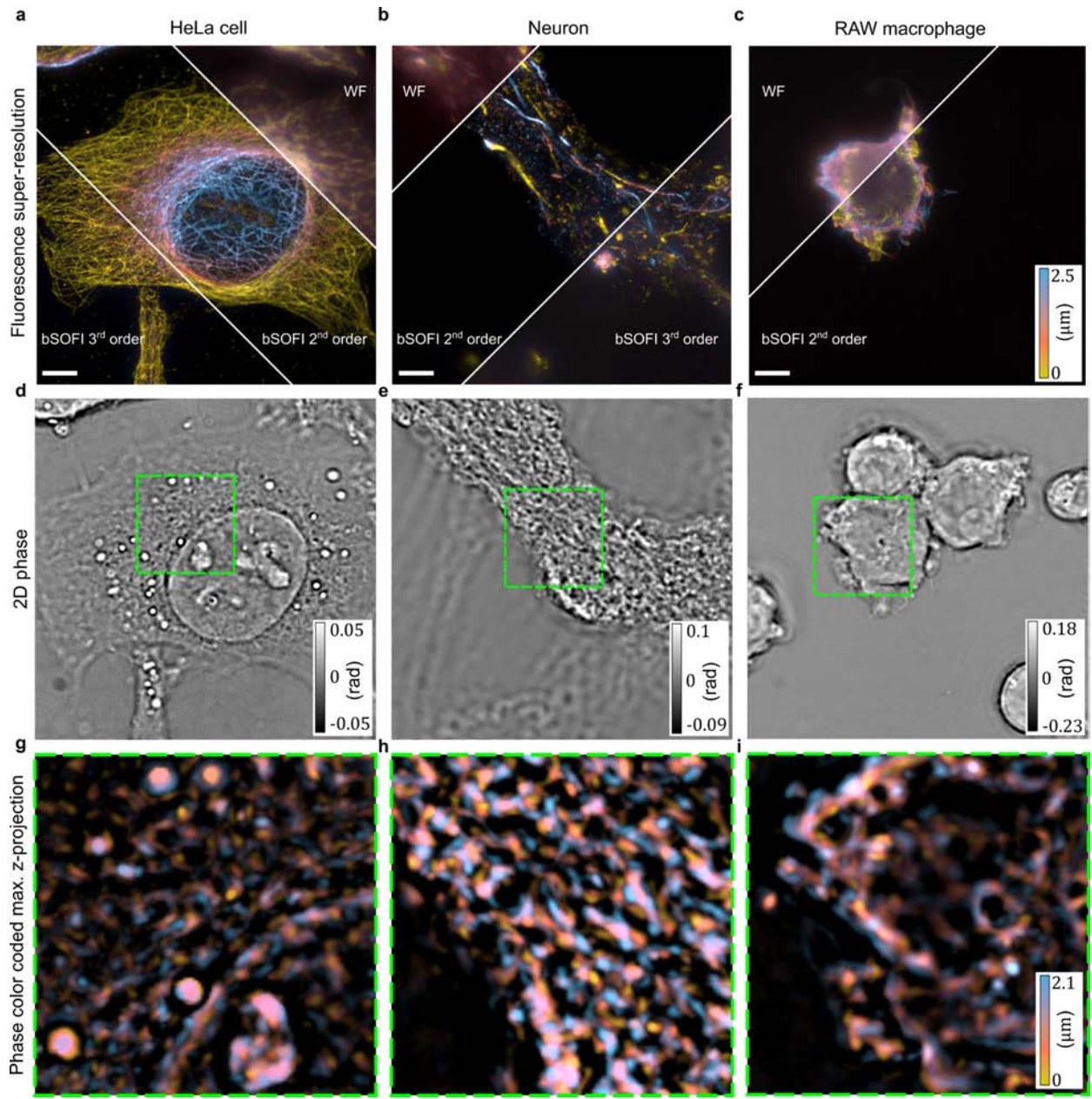

Figure 5: Multi-plane SOFI and phase imaging. SOFI maximum intensity projection, selected 2D phase image, and color coded maximum phase z-projection of the corresponding highlighted area (threshold T = 0 rad). (a), (d), (g) HeLa cell with Alexa Fluor 647 antibody-labelled microtubules, (b), (e), (h) mouse hippocampal primary neurites incubated with external α-synuclein fibrils and immunostained for newly formed alpha-synuclein aggregates Alexa647 and (c), (f), (i) live murine RAW 264.7 macrophages expressing Lifeact-Dreiklang. SOFI acquisition at 50 Hz of 5000 frames for Alexa647 and of 3000 frames for Dreiklang imaging. Subsequent phase imaging camera exposure time 20 ms, Scalebar 5 μm.

Next, we imaged live cells with phase tomography and SOFI. To do so, we transfected RAW 264.7 macrophages with a reversibly photoswitchable fluorescent protein construct, Lifeact-Dreiklang. This fluorescent fusion protein labels actin filaments in live cells[43], that are visualized in 2nd order bSOFI (Fig.

5 c). A visual comparison between fluorescence and phase imaging reveals that only one out of five cells in the field of view are expressing Lifeact-Dreiklang (Fig. 5c, f). The semi-adherent macrophages have a round shape with several thin, actin driven protrusions that emerge from the cellular membrane. In addition, we imaged the Vimentin network in HeLa cells and followed their dynamics by longer-term phase imaging (Vimentin-Dreiklang, see Supplementary Material Fig. S20 and Supplementary Movie M3). The diffusion of high-refractive index vesicles was monitored over 100 s by single particle tracking, providing an example of analysis of intracellular dynamics.

## Summary and discussion

In this study, we provide a novel concept for retrieving the phase information in all 3 spatial dimensions. The theory is based on the Helmholtz wave equation and is embedded in the framework of the generalized Wiener-Khintchine theorem[35]. We consider elastic light scattering using the first order Born approximation. An Ewald sphere construction reveals the attainable frequency support of the image spectrum. Measuring the interference of the forward weak scattering with the illumination enables decoding of the phase information along the z-direction. The theoretical effort goes in parallel with a fast and elegant algorithm comprising a masking operation to recover the 3D phase information from an acquired volumetric intensity stack. The common-path configuration integrated with white-light Koehler illumination and high detection NA provides stable and speckle-free high-resolution quantitative phase imaging. Our simulations confirmed and matched an experimental lateral and axial resolution of 350 nm and 560 nm. We demonstrated the quantitative nature of our method by assessing the topology of a nanometric phase object, confirmed by AFM. The method allows one to upgrade a classical brightfield microscope into a simple and reliable 3D phase microscope.

The experimental counterpart is based on an innovative multi-plane configuration, containing an image splitting prism for "volumetric" multiplexing, i.e. the simultaneous acquisition of 8 images originating from 8 conjugated object planes with an inter-plane distance of 350 nm. As shown, the advantage of high-speed 3D image acquisition entails as a compromise the loss of low spatial frequencies. Our Phase Retrieval Instrument with Super-resolution Microscopy (PRISM) combines fast 3D phase imaging with 3D fluorescence super-resolution microscopy for a unique 4D imaging modality.

This powerful and innovative concept opens the door to a wide range of applications, here demonstrated by imaging different cell samples. As shown, retrieving the 3D phase information at an acquisition rate up to 200 Hz responds to an ever growing demand for imaging fast dynamic cell processes. However, the gain in acquisition rate based on phase imaging is paired with a lack of specificity. Since our proposed MP configuration allows super-resolved fluorescence imaging based on super-resolution optical fluctuation imaging, fluorescence based specificity and high acquisition speed are realized in the same 3D platform. This multimodal and versatile microscope promises to fulfill the expectations for many novel applications and investigation in biology and life sciences.

**Methods**

*Microscope Setup*

All imaging was performed with a custom built microscope equipped with a temperature and $CO_2$ controlled incubator for live cell imaging[10]. For phase imaging we used a white-light Koehler illumination module. This optical system corresponds to a classical wide-field setup with an integrated low magnification telescope (see Fig. 3 and Supplementary Material Section 6).

For fluorescence imaging, a 120 mW, 405 nm laser (iBeam smart, Toptica), a 800 mW, 635 nm laser (MLL-III-635, Roithner Lasertechnik), and a 800 mW, 532 nm laser (MLL-FN-532, Roithner Lasertechnik) were combined for wide-field epi-fluorescence illumination and focused into the back focal plane of an Olympus UPLSAPO 60XW 1.2 NA objective. The fluorescence light was filtered using a combination of a dichroic mirror (zt405/488/532/640/730rpc, Chroma) and an emission filter adapted to the respective experiment (see Supplementary Material Table S2).

For phase imaging, we used the Koehler illumination module of a Zeiss Axiovert 100 M microscope equipped with a halogen lamp (spectrum see Supplementary Material Fig. S8). The detection path is arranged as a sequence of four 2-f configurations to provide image-object space telecentricity. The image splitter placed behind the last lens directs the light into 8 images, which are registered by two synchronized sCMOS cameras (ORCA Flash 4.0, Hamamatsu; back projected pixel size of 111 nm). For translating the sample, the microscope is equipped with piezoLEGS® stage (3-PT-60- F2,5/5) and Motion-Commander-Piezo controller (Nanos Instruments GmbH).

A more detailed description of the setup is given in the Supplementary Material Section 6 and Table S1.

### Cell Culture

HeLa cells, HEK293T cells, human fibroblasts and RAW 264.7 cells were cultured at 37 °C and 5 % $CO_2$ using DMEM high glucose with pyruvate (4.5 g $l^{-1}$ glucose, with GlutaMAX$^{TM}$ supplement, gibco®, Thermo Fisher Scientific or Roti®-CELL DMEM, Roth) supplemented with 10 % fetal bovine serum and 1 x penicillin-streptomycin (both gibco®, Thermo Fisher Scientific).

### RAW 264.7 Stimulation

RAW 264.7 cells were stimulated with 100 ng $ml^{-1}$ lipopolysaccharides from *E. coli* O111:B4 in DMEM (see cell culture) for about 4 h at 37 °C and 5 % $CO_2$.

### Cell fixation for phase and fluorescence imaging

Cells were seeded in Lab-tek® II chambered cover slides (nunc) or in FluoroDish Sterile Culture Dish 35 mm, 23 mm well (World Precision Instruments) 1-2 days before fixation in DMEM (see cell culture) or DMEM high glucose w/o phenol red (4.5 g $l^{-1}$ glucose) supplemented with 4 mM L-gluthamine, 10 % fetal bovine serum and 1x penicillin-streptomycin (all gibco®, Thermo Fisher Scientific).

Cells were washed twice in pre-warmed microtubule stabilizing buffer (MTSB: 100 mM PIPES pH 6.8, 2mM $MgCl_2$, 5 mM EGTA), followed by application of pre-warmed fixation buffer (MTSB with 3.7 % paraformaldehyde (PFA) and 0.2 % Triton X-100 for microtubule immunostaining) for 15 min at room temperature (RT). Cells were then washed three times for 5 min each with 1 x PBS and stored in 50 % glycerol in 1 x PBS at 4 °C until phase imaging or the immunostaining protocol was continued to prepare samples for fluorescence imaging.

### Immunostaining HeLa Cells

Fixed and permeabilized cells were blocked with 3 % BSA in 1 x PBS and 0.05 % Triton X-100 for 60 min at RT or overnight at 4 °C. The blocked samples were immediately incubated with primary anti-tubulin antibody (0.01 mg $ml^{-1}$ DM1a mouse monoclonal, Abcam) in antibody incubation buffer for 60 min at RT (AIB: 1 % BSA in 1 x PBS and 0.05 % Triton X-100). Cells were then washed three times for 5 min each with AIB, followed by incubation with donkey anti-mouse-Alexa Fluor 647 antibody (0.01 mg $ml^{-1}$ Invitrogen) for 60 min at RT. All subsequent steps were performed in the dark. Cells were again washed three times for 5 min each with AIB, incubated for 15 min post-fixation with 2 % PFA in 1 x PBS followed by three 5 min washes with PBS. Cells were imaged immediately or stored in 50 % glycerol in 1 x PBS at 4 °C until imaging.

### Transfection and Plasmid Construction

Cells were transfected using FuGENE® 6 transfection reagent (Promega) at about 80 % confluency in FluoroDish Sterile Culture Dishes 35 mm, 23 mm well (World Precision Instruments) according to manufacturer's protocols. For each dish with HeLa cells, 6 µl FuGENE® 6 were mixed with 92 µl OptiMEM Reduced-Serum Medium (Life Technologies) and incubated for 5 min before the addition of 2 ng of the plasmid pMD-Vim-Dreiklang[44]. For each dish with RAW 264.7 cells, 9.6 µl FuGENE® 6 were mixed with 125.2 µl OptiMEM Reduced-Serum Medium (Life Technologies) and incubated for 5 min before the addition of 2.8 ng of the plasmid pCDNA3- Lifeact-Dreiklang. After incubation for 30 min at RT, 100 µl and 125 µl of the solution was carefully distributed over the cells that were supplied with fresh medium, respectively. Cells were then returned to the incubator and left overnight before imaging.

Lifeact-Dreiklang was constructed as follows: The Dreiklang sequence (Aberrior) was amplified by PCR and inserted in the plasmid pCDNA3 (Invitrogen) using PmeI and AgeI restriction sites. The Lifeact sequence was synthesized (Microsynth) according to Riedl et al.[43], annealed and inserted in front of Dreiklang using KpnI and BstBI restriction sites.

### Primary Neuron Culture Preparation

Pregnant female C57BL/6J Rcc Hsd were obtained from Harlan Laboratories (France) and were housed according to the Swiss legislation and the European Community Council directive (86/609/EEC). Primary Hippocampal cultures were prepared from mice brains from P0 pups as previously described[45]. Briefly, the hippocampi were isolated stereoscopically and dissociated by trituration in medium containing papain (20 U mL[-1], Sigma-Aldrich, Switzerland). The neurons were plated on FluoroDish Sterile Culture Dish 35 mm, 23 mm well (World Precision Instruments), previously coated with poly-L-lysine 0.1 % w/v in water, and neurons were cultured in Neurobasal medium containing B27 supplement (Life Technologies), L-glutamine and penicillin/streptomycin (100 U mL[-1], Life Technologies).

### Treatment of Hippocampal Neurons with α-synuclein (α-syn) Pre-formed Fibrils

After 5 days *in vitro*, the hippocampal neurons were treated with α-syn pre-formed fibrils (PFFs) prepared and characterized as previously described[46]. Briefly, α-syn PFFs were diluted in the neuronal cell culture media at a final concentration of 70 nM and added to the hippocampal neurons for ten days as described by Lee's group[47,48]. The extracellular α-syn PFFs, once internalized by the hippocampal primary neurons, recruit endogenous α-syn which leads to the formation of new intracellular aggregates hyperphosphorylated on the S129 residue of α-syn. These newly formed aggregates can be detected by immunocytochemistry ten days post-treatment using an antibody that specifically stains α-syn phosphorylated on S129 (pS129)[47,48].

### *Staining of Neurons*

The Hippocampal primary neurons were washed twice with PBS (Life Technologies) and were then fixed in 4 % paraformaldehyde for 20 minutes at RT. After two washed in PBS, neurons were incubated in 3 % bovine serum albumin (BSA) in 0.1 % Triton X-100 PBS (PBS-T) for 30 min at RT.  Hippocampal primary neurons were then incubated with primary antibodies [chicken anti-MAP2 (Abcam) to specifically stain the neurons, mouse anti-α-syn (SYN-1, BD) to detect the total level of α-syn and rabbit anti-pS129-α-syn (MJFR-13, Abcam) to detect the newly formed aggregates] for 2 hours at RT. The cells were rinsed 5 times in PBS-T and subsequently incubated respectively with the secondary donkey anti-chicken Alexa488 (Jackson ImmunoResearch), anti-mouse Alexa594 and anti-rabbit Alexa647 (Life Technologies) at a dilution of 1/800 in PBS-T and DAPI at 2 μg/ml (Life Technologies). The cells were washed five times in PBS-T and twice in PBS before being imaged.

### *Imaging Media/Buffer*

Phase imaging of fixed samples was performed in 1 x PBS.

The Alexa647 immunostained samples for SOFI were imagined in a 50 mM Tris-Hcl pH 8.0, 10 mM NaCl buffer containing an enzymatic oxygen scavenging system (2.5 mM protocatechuic acid (PCA) and 50 nM Protocatechuate- 3,4-Dioxygenase from Pseudomonas Sp. (PCD) with >3 Units m $g^{-1}$)[49] and a thiol (2-Mercaptoethylamine). The thiol and a stock solution of 100 mM PCA in water, pH adjusted to 9.0 with NaOH, were prepared on the measurement day and PCD was aliquoted at a concentration of 10 μM in storage buffer (100 mM Tris-HCl pH 8.0, 50 % glycerol, 50 mM KCl, 1 mM EDTA) at -20 °C.

### *Live cell imaging*

Label-free imaging was performed in DMEM or DMEM w/o phenol red at 35 °C and 5 % $CO_2$ (for details see Cell Culture). For living cells with Dreiklang plasmids, the medium was exchanged to antibleaching medium DMEM$^{gfp}$-2 supplemented with rutin (Evrogen) at a final concentration of 20 mg $l^{-1}$ approx. 15-30 min before imaging.

### *Chemicals*

Unless noted otherwise, all chemicals were purchased at Sigma-Aldrich.

## Acknowledgements

We thank P. Sandoz for construction of Lifeact-Dreiklang (VDG, EPFL), the LSBG (EPFL) for providing RAW 264.7 cells and M. Ricchetti from Institute Pasteur for human fibroblast cells. We are grateful to M. Sison for cell culture advice and assistance. We thank O. Peric (LBNI, EPFL) for providing the technical sample and performing the AFM measurement. We acknowledge A. Radenovic and A. Nahas for support and discussion and G. M. Hagen for proofreading of the manuscript. This project has received funding from the European Union's Horizon 2020 research and innovation program under the Marie Skłodowska-Curie Grant Agreement No. [750528]. The research was supported by the Swiss National Science Foundation (SNSF) under grant 200020_159945/1. T.L. acknowledges the support from the Horizon 2020 Framework Programme of the European Union via grant 686271.

## Author Contributions

T.L. and A.D. initiated the project and wrote the theory/modeling. A.D. developed the phase retrieval algorithm and simulations. K.S.G., A.D., T.L. and A.S. designed the experiments. K.S.G. prepared and performed the experiments. A.-L. M.-M. prepared the neuron samples. A.D., K.S.G. and T.Lu. analyzed the data. M.L., S.G. and T.L. designed the optical system including the image splitting prism. S.G. and A.S. built the microscope setup. E.B., A.B. and H.A.L. provided research advice. A.D., K.S.G. and T.L. wrote the manuscript with contributions from all authors.

## Competing Financial Interest Statement

The authors declare no competing financial interests.

# Combined Multi-Plane Tomographic Phase Retrieval and Super-Resolution Optical Fluctuation Imaging for 4D Cell Microscopy - Supplementary Material

## Contents

**Combined Multi-Plane Tomographic Phase Retrieval and Super-Resolution Optical Fluctuation Imaging for 4D Cell Microscopy - Supplementary Material**

# 1 Polychromatic light scattering

## 1.1 Monochromatic scattering[2]

Following the seminal work of Born & Wolf[1], we consider a monochromatic scalar electromagnetic field $U(\boldsymbol{x})$ at frequency $\omega$ (assuming a slowly varying refractive index $n(\boldsymbol{x})$). This electromagnetic field has to satisfy the Helmholtz equation

$$(\nabla^2 + k_0^2 n^2(\boldsymbol{x}))U(\boldsymbol{x}) = 0, \tag{S1}$$

where $\boldsymbol{x} = (x, y, z)$ is the spatial coordinate, $\nabla^2 = \frac{\partial^2}{\partial x^2} + \frac{\partial^2}{\partial y^2} + \frac{\partial^2}{\partial z^2}$ is the Laplacian operator, $k_0 = \frac{\omega}{c} = \frac{2\pi}{\lambda}$ the wavenumber of the electromagnetic field in vacuum and $n(\boldsymbol{x})$ being the spatial refractive index distribution of the sample.

Equation (S1) cannot be solved in general due to the spatially varying refractive index. We therefore take the index of refraction to vary around the mean refractive index $\bar{n}$. Equation (S1) can now be rewritten as

$$(\nabla^2 + k_0^2\,\bar{n}^2\,)U(\boldsymbol{x}) = -F(\boldsymbol{x})U(\boldsymbol{x}), \tag{S2}$$

where $F(\boldsymbol{x}) = k_0^2(n^2(\boldsymbol{x}) - \bar{n}^2)$ is the scattering potential.

The left-hand side of equation (S2) is the homogeneous wave equation, whereas the right hand side represents a source term containing the scattering potential $F(\boldsymbol{x})$.

Outside of the scattering volume, the total field is written as a superposition of the incident field and the scattered field

$$U(\boldsymbol{x}) = U_i(\boldsymbol{x}) + U_s(\boldsymbol{x}). \tag{S3}$$

Using the homogeneous Helmholtz equation for the incident field $U_i(\boldsymbol{x})$, we obtain

$$(\nabla^2 + \bar{n}^2 k_0^2)U_i(\boldsymbol{x}) = 0. \tag{S4}$$

By combining equations (S2) and (S3) and using (S4), we write for the scattered field $U_s(\boldsymbol{x})$, which satisfies the inhomogeneous Helmholtz equation

---

[2] Throughout all the calculations we use the same notation for the intensity and its Fourier transformed spectrum. The distinction is made by explicitly noting the variable $\boldsymbol{x}$ or $\boldsymbol{g}$ and the bold font is reserved for the spatial coordinates in three dimensions, i.e. $\boldsymbol{x} = (x, y, z)$ and $\boldsymbol{g} = (g_\perp, g_z)$, with $g_\perp$ shorthand for $(g_x, g_y)$. The wavenumber, i.e. the magnitude of the wave vector, is denoted as $k_0 = \sqrt{{k_\perp}^2 + {k_z}^2} = \frac{2\pi}{\lambda} = \frac{\omega}{c}$.

# Combined Multi-Plane Tomographic Phase Retrieval and Super-Resolution Optical Fluctuation Imaging for 4D Cell Microscopy - Supplementary Material

$$[\nabla^2 + k_0^2\, \bar{n}^2] U_s(\boldsymbol{x}) = -F(\boldsymbol{x}) U(\boldsymbol{x}). \tag{S5}$$

Using a Green's function ansatz[1] we write

$$[\nabla^2 + k_0^2\, \bar{n}^2] G(\boldsymbol{x} - \boldsymbol{x}') = -\delta(\boldsymbol{x} - \boldsymbol{x}'). \tag{S6}$$

where $\delta(\boldsymbol{x} - \boldsymbol{x}')$ is the 3D Dirac function.

Approximating the far-field response of a point scatterer as a spherical wave, we have

$$G(\boldsymbol{x} - \boldsymbol{x}') = \frac{e^{\mathrm{i} k_0 |\boldsymbol{x} - \boldsymbol{x}'|}}{|\boldsymbol{x} - \boldsymbol{x}'|}. \tag{S7}$$

Because the scattered field is much weaker than the incident field (first-order Born approximation), we neglect the scattered field in comparison to the dominant incident field. We obtain outside of the scattering volume $V_s$ for the scattering field

$$U_s(\boldsymbol{x}) = \int_{V_s} G(\boldsymbol{x} - \boldsymbol{x}') F(\boldsymbol{x}') U_i(\boldsymbol{x}') \mathrm{d}^3 x'. \tag{S8}$$

This represents a 3D convolution of the source term $F(\boldsymbol{x}') U_i(\boldsymbol{x}')$ with the Green's function. The Green's function (equation (S7)) can be expressed by a series expansion (Weyl's expansion of a spherical wave in terms of plane waves) in the lateral coordinate x and y as[2,3]

$$G(\boldsymbol{x} - \boldsymbol{x}') = \frac{\mathrm{i}}{2\pi k_z} \iint_{-\infty}^{\infty} e^{\mathrm{i}\left(k_x(x-x') + k_y(y-y') + k_z(z-z')\right)} \mathrm{d}k_x \mathrm{d}k_y. \tag{S9}$$

Inserting equation (S9) into (S8), we obtain for $U_s(\boldsymbol{k}; z)$ in Fourier space

$$U_s(\boldsymbol{k}; z) = \frac{\mathrm{i}}{k_z} e^{\mathrm{i} k_z z} [F(\boldsymbol{g}) \otimes U_i(\boldsymbol{g})], \tag{S10}$$

where $\boldsymbol{g}$ is the object spatial frequency space and $\boldsymbol{k}$ is the field spatial frequency space, $\otimes$ is the convolution operator, $e^{\mathrm{i} k_z z}$ represents the propagation of the wave and $k_z = \bar{n} k_0 \cos(\theta_{det})$ represents the projection of the scattering vector $\boldsymbol{k}$ onto the optical axis (where we adopt the notation of Singer et al.[4]).

Assuming a monochromatic plane wave (in Fourier space) with an amplitude $A$, propagating along the direction $\boldsymbol{k}_i$ for the incident field $U_i(\boldsymbol{g}; \boldsymbol{k}_i)$, expressed as

$$U_i(\boldsymbol{g}; \boldsymbol{k}_i) = A\delta(\boldsymbol{g} - \boldsymbol{k}_i), \tag{S11}$$

and neglecting the propagation term ($z = 0$), we obtain a general expression where the scattered field appears as an interaction of the illumination field and the scattering potential, given as

$$U_s(\boldsymbol{k}; \boldsymbol{k}_i) = \mathrm{i} A \frac{F(\boldsymbol{k} - \boldsymbol{k}_i)}{k_z}. \tag{S12}$$

The object spectrum or scattering potential is interrogated by the illumination field, represented by its wavevector $\boldsymbol{k}_i$. The scattering event results in a plane wave with an amplitude $U_s(\boldsymbol{k}; \boldsymbol{k}_i)$,

**Combined Multi-Plane Tomographic Phase Retrieval and Super-Resolution Optical Fluctuation Imaging for 4D Cell Microscopy - Supplementary Material**

propagating along the direction $\boldsymbol{k}$.

## 1.2    Coherent transfer function

The microscope is modeled as a telecentric, diffraction limited optical imaging system. The microscope is fully characterized by the source spectrum and the illumination and detection Numerical Aperture ($NA_{ill}, NA_{det}$). Due to the limited bandwidth of our detection system (sCMOS camera), the intensity is described by a temporal average (denoted <>) of the interference between the scattered and un-scattered field and integrated over the source spectrum ($\omega$) the angular spectra ($\boldsymbol{k_i}, \boldsymbol{k}$).

$$\text{I}(\boldsymbol{x}) = <\left| \iint_{\omega, \boldsymbol{k_i}, \boldsymbol{k}} U_i(\boldsymbol{x}; \omega, \boldsymbol{k_i}) + U_s(\boldsymbol{x}; \omega, \boldsymbol{k_i}, \boldsymbol{k}) \, \mathrm{d}\boldsymbol{k}\mathrm{d}\boldsymbol{k_i}\mathrm{d}\omega \right|^2 > . \tag{S13}$$

Developing the equation S13, we decompose the intensity into a sum of mutual intensities as

$$\text{I}(\boldsymbol{x}) = < \int_{\omega', \boldsymbol{k_i'}} \int_{\omega'', \boldsymbol{k_i''}} U_i(\boldsymbol{x}; \omega', \boldsymbol{k_i'}) \, U_i^*(\boldsymbol{x}; \omega'', \boldsymbol{k_i''}) \, \mathrm{d}\omega' \mathrm{d}\boldsymbol{k_i'} \mathrm{d}\omega'' \mathrm{d}\boldsymbol{k_i''} > + \tag{S14}$$

$$< \int_{\omega', \boldsymbol{k_i'}, \boldsymbol{k'}} \int_{\omega'', \boldsymbol{k_i''}} U_s(\boldsymbol{x}; \omega', \boldsymbol{k_i'}, \boldsymbol{k'}) U_i^*(\boldsymbol{x}; \omega'', \boldsymbol{k_i''}) \, \mathrm{d}\omega' \mathrm{d}\boldsymbol{k_i'} \mathrm{d}\boldsymbol{k'} \mathrm{d}\omega'' \mathrm{d}\boldsymbol{k_i''} > + c.c.$$

where $c.c.$ denotes the complex conjugate term $U_s U_i$. We neglected the weak $U_s U_s^*$ contribution due to the weak scattering approximation.

Using the generalized Wiener-Khintchine theorem[5], the mutual intensity of two fields is only non-zero for $\omega' = \omega''$. Following the work of N. Streibl[6], which applies for a telecentric configuration and a Koehler illumination, the mutual intensity is non-zero only for $\boldsymbol{k_i'} = \boldsymbol{k_i''}$. We then obtain

$$\text{I}(\boldsymbol{x}) = \int_{\omega, \boldsymbol{k_i}} U_i(\boldsymbol{x}; \omega, \boldsymbol{k_i}) U_i^*(\boldsymbol{x}; \omega, \boldsymbol{k_i}) \, \mathrm{d}\omega \mathrm{d}\boldsymbol{k_i} + \tag{S15}$$

$$\int_{\omega, \boldsymbol{k_i}, \boldsymbol{k}} U_s(\boldsymbol{x}; \omega, \boldsymbol{k_i}, \boldsymbol{k}) U_i^*(\boldsymbol{x}; \omega, \boldsymbol{k_i}) \mathrm{d}\omega \mathrm{d}\boldsymbol{k_i} \mathrm{d}\boldsymbol{k} + c.c.$$

The first term represents the unscattered field and the second term the mutual interference between the illumination and scattered fields.

The product between the scattered and un-scattered field contains all information of the object. Following equation (S11) and (S12), we obtain in the object Fourier space

$$U_s(\boldsymbol{g}; \omega, \boldsymbol{k}) \otimes U_i^*(\boldsymbol{g}; \omega, \boldsymbol{k_i}) =$$

$$\mathrm{i}A(\omega) \frac{F(\boldsymbol{g}; \omega) \delta(\boldsymbol{g} - \boldsymbol{k})}{k_z} \otimes A^*(\omega) \delta(\boldsymbol{g} + \boldsymbol{k_i}) = \mathrm{i}S(\omega) \frac{F(\boldsymbol{g}; \omega)}{k_z} \delta(\boldsymbol{g} - \boldsymbol{K}), \tag{S16}$$

where $\boldsymbol{K} = \boldsymbol{k} - \boldsymbol{k_i}$ (Laue equation) and $AA^* = S(\omega)$ the intensity of the light source. This expression is illustrated in Fig. S1a by the corresponding Ewald sphere construction taking into account the elastic light scattering. For a given frequency $\omega$, illumination and scattering vector $\boldsymbol{k_i}$ and $\boldsymbol{k}$, the scattering potential $F(\boldsymbol{g}; \omega)$ at the spatial frequency $\boldsymbol{K}$ is interrogated with an amplitude $\frac{S(\omega)}{k_z}$.

Finally the interference term i.e. the cross-spectral density $\Gamma(\boldsymbol{g})$ is expressed

## Combined Multi-Plane Tomographic Phase Retrieval and Super-Resolution Optical Fluctuation Imaging for 4D Cell Microscopy - Supplementary Material

$$\Gamma(\boldsymbol{g}) = \int_{\omega, \boldsymbol{k_i}, \boldsymbol{k}} \mathrm{i} \frac{S(\omega)F(\boldsymbol{g};\omega)\delta(\boldsymbol{g}-\boldsymbol{K})}{k_z} \mathrm{d}\boldsymbol{k_i}\mathrm{d}\boldsymbol{k}\mathrm{d}\omega \qquad (S17)$$

where we integrate over the source spectra and the angular spectra (illumination and scattering).

Due to the limited bandwidth of our illumination, we neglect the dispersion of the scattering potential. This results in a linear relationship for the cross-spectral density.

$$\Gamma(\boldsymbol{g}) = \mathrm{i}F(\boldsymbol{g})H(\boldsymbol{g}) \qquad (S18)$$

with the polychromatic coherent transfer function $H(\boldsymbol{g})$,

$$\mathrm{H}(\boldsymbol{g}) = \int_{\omega, \boldsymbol{k_i}, \boldsymbol{k}} \frac{S(\omega)}{k_z} \delta(\boldsymbol{g}-\boldsymbol{K}) \mathrm{d}\boldsymbol{k_i}\mathrm{d}\boldsymbol{k}\mathrm{d}\omega. \qquad (S19)$$

Each combination of illumination and detection modes (frequency ($\omega$) dependent) interrogates a different point of the object's spatial frequency content. The final transfer function is then given by a linear superposition of all contributions.

We obtained this result based on the Helmholtz equation using the first order Born approximation for describing the scattering field as an interaction between the illumination field and a weakly scattering object. The polychromatic illumination has been embedded in a generalized Wiener-Khintchine formalism. The spatial coherence is taken into account by a mutual intensity consideration following the seminal work of N. Streibl[6]. This work on 3D imaging is based on a telecentric configuration containing a Koehler illumination matching all experimental elements of our setup. As a main result of this analysis, the scattering potential $F(\boldsymbol{g})$ is low pass filtered by the imaging system as described by the Coherent Transfer Function (CTF) $H(\boldsymbol{g})$.

A key assumption we made to derive this linear model (equation (S18)) is the weak scattering approximation, i.e. only single scattering events contribute to the measured signal. This assumption is valid for example when imaging single layer of cells. Imaging thicker samples requires a modified theory taking into account multiple scattering events[1].

The complex ingredients of the theoretical analysis are illustrated in an Ewald sphere representation (Fig. S1b). For each wavelength, the Ewald sphere shows the frequency support which corresponds to an axially shifted sphere cap of radius $\bar{n}k_0$ with a lateral extent given by the product of the rescaled wave number $k_0$ with the detection numerical aperture ($g_{\perp max} = \bar{n}k_0\,NA_{det}$) and an axial extent limited to[7] $g_{z,\max}(g_\perp, NA_{det}) = \bar{n}g_\perp \left(1 - \sqrt{1 - \frac{NA_{det}^2}{\bar{n}^2}}\right)$. Each of these wavelength-dependent sphere caps are summed up to build the support of the polychromatic system transfer function, where the weights are given according to equation (S19).

Taking into account the full angular spectrum of the illumination, we integrate over all illumination $\boldsymbol{k_i}$. The resulting CTF $H(\boldsymbol{g})$ is shown in Fig. S1c.

**Combined Multi-Plane Tomographic Phase Retrieval and Super-Resolution Optical Fluctuation Imaging for 4D Cell Microscopy - Supplementary Material**

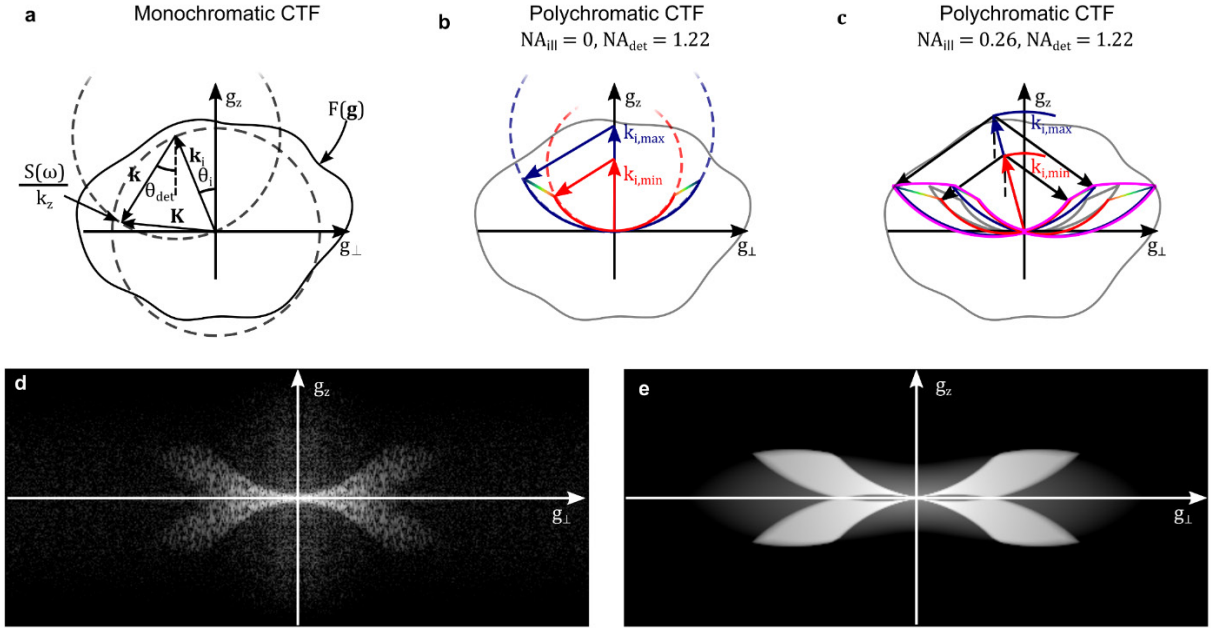

Figure S1 Construction of the polychromatic coherent transfer function. (a) Illustration of elastic light scattering (b) Frequency support of the polychromatic CTF with plane wave illumination, (c) Frequency support of the polychromatic CTF with Koehler illumination, (d) Logarithmic absolute valued 3D-FFT of experimental intensity stack, acquired by translating a sample of fixed hippocampal primary neurons in steps of 200 nm ($\approx$50 µm x 50 µm x 40 µm), showing the system transfer function and its mirrored complex conjugate, (e) Logarithmic absolute valued 3D-FFT of simulated 3D intensity stack based on the proposed model ($NA_{ill} = 0.26, NA_{det} = 1.15$ source spectrum (Table S1)).

In complement to the theoretical analysis and the Ewald sphere representation, we calculate (Fig. S1d) the experimental 3D-fast Fourier Transform (FFT) of a large intensity stack, containing the transfer function $H(\boldsymbol{g})$ and its mirrored complex conjugate. Fig. S1e shows the simulated 3D-FFT of an equally large intensity stack, based on the experimental source spectrum, the illumination and detection NA.

## 2 Retrieving the complex 3D cross-spectral density $\boldsymbol{\Gamma(g)}$

To recover the 3D cross-spectral density and its corresponding phase, only a 3D brightfield image stack is required. The different planes can be acquired sequentially by a z-scan or in a multi-plane configuration (see Fig. S5), which provides camera rate 3D phase tomographic measurements. For each plane $z_p$ we detect the interference of the incident field with the scattered field. Using equations (S15, S17), we write

$$I(\boldsymbol{x_p}) = I_{DC} + \Gamma(\boldsymbol{x_p}) + \Gamma^*(\boldsymbol{x_p}) \,, \tag{S20}$$

where we used the short-hand notation $\boldsymbol{x_p} = (x, y, z_p)$.

Taking the 3D Fourier transform of this equation, we obtain

$$I(\boldsymbol{g}) = I_{DC}\delta(\boldsymbol{g}) + \Gamma(\boldsymbol{g}) + \Gamma^*(-\boldsymbol{g}). \tag{S21}$$

# Combined Multi-Plane Tomographic Phase Retrieval and Super-Resolution Optical Fluctuation Imaging for 4D Cell Microscopy - Supplementary Material

The intensity spectrum can be decomposed into a DC-part $I_{DC}\delta(\boldsymbol{g})$ and an AC part, composed of two symmetric and conjugate cross-spectral densities. For small illumination NA, the cross-spectral density essentially expands into the subspace $g_z \geq 0$ (see Figure S1c,[3]). Both mirror-contributions $\Gamma(\boldsymbol{g})$ and $\Gamma^*(-\boldsymbol{g})$ of the scattering potential are largely dissociated.

Filtering the complex field $I(\boldsymbol{g})$

$$\Gamma_+(\boldsymbol{g}) = I(\boldsymbol{g})K(g_z) \tag{S22}$$

with a cutoff filter $K(g_z)$

$$K(g_z) = \begin{cases} 1 & if \ g_z > g_{z,c} \\ 0 & else \end{cases} \tag{S23}$$

suppresses both the conjugated cross-spectral density $\Gamma^*(-\boldsymbol{g})$ and the low frequency overlap.

This filtering condition entails the analyticity of the cross-spectral density along the axial direction i.e. the real and imaginary part of $\Gamma(x)$ form a Hilbert transform pair (Titchmarsh theorem)[8]. Therefore, the amplitude and the phase are simply two alternative representations of the filtered scattering potential, where the amplitude appears as the imaginary part of transfer function $H(\boldsymbol{g})$ and the phase as the real part of $H(\boldsymbol{g})$ (see S18). In other words, a point scatterer appears in intensity as an axial phase shift with no contrast in focus while the phase appears as a Gaussian with maximum contrast (see Fig. S3a).

The threshold $g_{z,c}$ is determined by the illumination NA (Fig. S1c). This translates directly to the suppression of the overlap of $\Gamma(\boldsymbol{g})$ and its mirrored complex conjugate $\Gamma^*(-\boldsymbol{g})$. The filter $K(g_z)$ acts as a high pass filter and removes this low frequency overlap.

We obtain for the cutoff

$$g_{z,c} = \bar{n}k_{0,max}\left(1 - \sqrt{1 - NA_{ill}^2}\right), \tag{S24}$$

where $k_{0,max} = \frac{2\pi}{\lambda_{min}}$. For our case with a $NA_{ill} \approx 0.26$, we suppress all frequencies $g_z < g_{z,c} \approx 0.4\ [\mu m^{-1}]$. As we are interested in the scattering potential of a cell and its subcellular structure, this cutoff has no major impact on the image quality.

Retrieving $\Gamma_+(\boldsymbol{g})$ (equation (S22)) leads to the phase difference between the scattered field and the reference field as $\Delta\varphi(\boldsymbol{x}) = angle\big(\Gamma_+(\boldsymbol{x})\big)$. However, the quantitative phase associated to the image field is given, according to [9], as

$$\varphi(\boldsymbol{x}) = \tan^{-1}\left(\frac{\beta(\boldsymbol{x})\sin(\Delta\varphi(\boldsymbol{x}))}{1 + \beta(\boldsymbol{x})\cos(\Delta\varphi(\boldsymbol{x}))}\right), \tag{S25}$$

where $\beta = \frac{|U_s(\boldsymbol{x})|}{|U_i(\boldsymbol{x})|}$ and $\Delta\varphi(\boldsymbol{x}) = angle\big(\Gamma_+(\boldsymbol{x})\big)$. After expansion of this fraction by $\frac{|U_i|^2}{|U_i|^2}$, we write

$$\varphi(\boldsymbol{x}) = \tan^{-1}\left(\frac{|U_i||U_s|\sin(\Delta\varphi)}{|U_i|^2 + |U_i||U_s|\cos(\Delta\varphi)}\right), \tag{S26}$$

# Combined Multi-Plane Tomographic Phase Retrieval and Super-Resolution Optical Fluctuation Imaging for 4D Cell Microscopy - Supplementary Material

and identify the imaginary and real part of the cross-spectral density as $|U_i||U_s|\sin(\Delta\varphi) = Im(\Gamma_+(\boldsymbol{x}))$, $|U_i||U_s|\cos(\Delta\varphi) = Re(\Gamma_+(\boldsymbol{x}))$, as well as the illumination intensity $|U_i|^2 = I_0$.

We obtain for the quantitative phase in relation to $\Gamma_+(\boldsymbol{x})$

$$\varphi(\boldsymbol{x}) = \tan^{-1}\left(\frac{Im(\Gamma_+(\boldsymbol{x}))}{I_{DC} + Re(\Gamma_+(\boldsymbol{x}))}\right). \tag{S27}$$

## 2.1 Quantitative phase algorithm

The phase retrieval algorithm (see Fig. S2) can be summarized as follows

1. Data acquisition.
2. Signal mirroring along the axial direction to avoid boundary effects and 3D Fourier transform of the 3D stack.
3. Application of mask K for removing all $g_z \leq g_{z,c}$ contributions of the Fourier spectrum and Fourier denoising using a CTF-shaped mask.
4. Inverse Fourier transform for retrieving the cross spectral density in real space. Reconstruct the image field and calculate the 3D quantitative phase.

**STEP 1: Data aquisition and multi-plane coregistration**

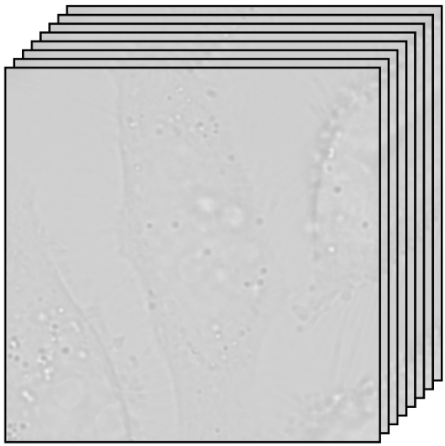

**STEP 2: Signal mirroring and 3D FFT**

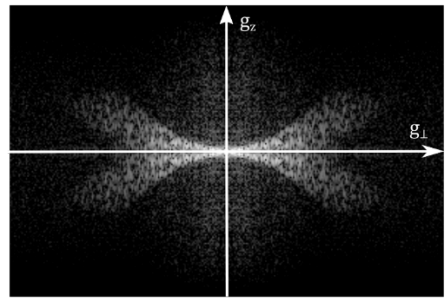

**STEP 3: Fourier filtering and denoising**

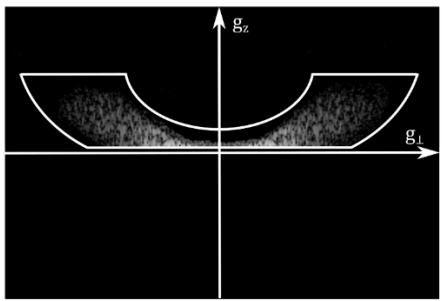

**STEP 4: i-FFT and phase calculation**

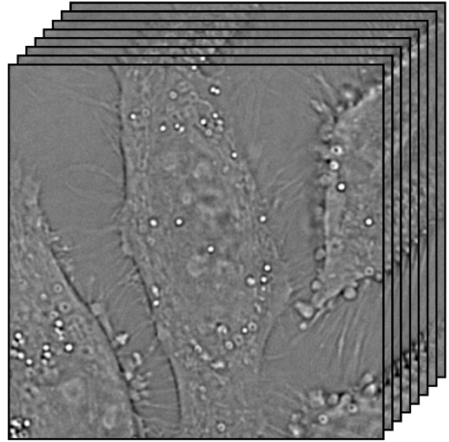

Figure S2: Workflow of the phase retrieval algorithm. In STEP 1 and 4, 8-plane image stacks of fixed HeLa Cell are shown. In STEP 2 and 3, the data is displayed using a logarithmic scale.

**Combined Multi-Plane Tomographic Phase Retrieval and Super-Resolution Optical Fluctuation Imaging for 4D Cell Microscopy - Supplementary Material**

## 3    Simulation and validation

### 3.1    Calculating the partially coherent system transfer function $H(g)$

For the illumination k-spectrum (limited by $NA_{ill}$), the corresponding Ewald sphere cap (limited by $NA_{det}$) is weighted according to the polychromatic spectrum of the source (Fig. S6) and projected on the reciprocal Fourier space (according to the specified field of view). These k-supports (Ewald sphere caps) are added to form the full polychromatic k-support. The calculated 2D CTF (basically a convolution of the illumination aperture and the detection aperture) is then mapped onto the 3D frequency support, taking into account the symmetry properties of the CTF (weighing according to equation (S16)), as shown in Fig. S1d.

### 3.2    3D image formation simulation

We define a point scatterer in real space by a Dirac function. The corresponding scattering potential is calculated and convolved with the previously established system transfer function $H(g)$ resulting in the complex 3D cross-spectral density $\Gamma(x)$. The 3D image intensity is given by the absolute squared interference of the incident field with the scattered field (equation (S18)). All calculations have been performed using Matlab (R2016a).

### 3.3    Experiment vs simulation

We imaged 200 nm polystyrene beads sparsely distributed in agarose. Due to their size ($< \lambda/2$), we used the beads as an approximation of a point scatterer.

Fig S3a displays an axial cross-section of the computed and experimental 3D image (averaged over 15 individual measurements), showing an almost perfect match between experimental and simulated images. Their corresponding calculated phase underlines the validity of our model and is further demonstrated in lateral and axial line plots (Fig. S3b). Fig. S3c shows the color-coded maximal z-projection of the full experimental recovered phase stack. The orthogonal slice 1-2 shows the optical sectioning for our 3D phase imaging.

**Combined Multi-Plane Tomographic Phase Retrieval and Super-Resolution Optical Fluctuation Imaging for 4D Cell Microscopy - Supplementary Material**

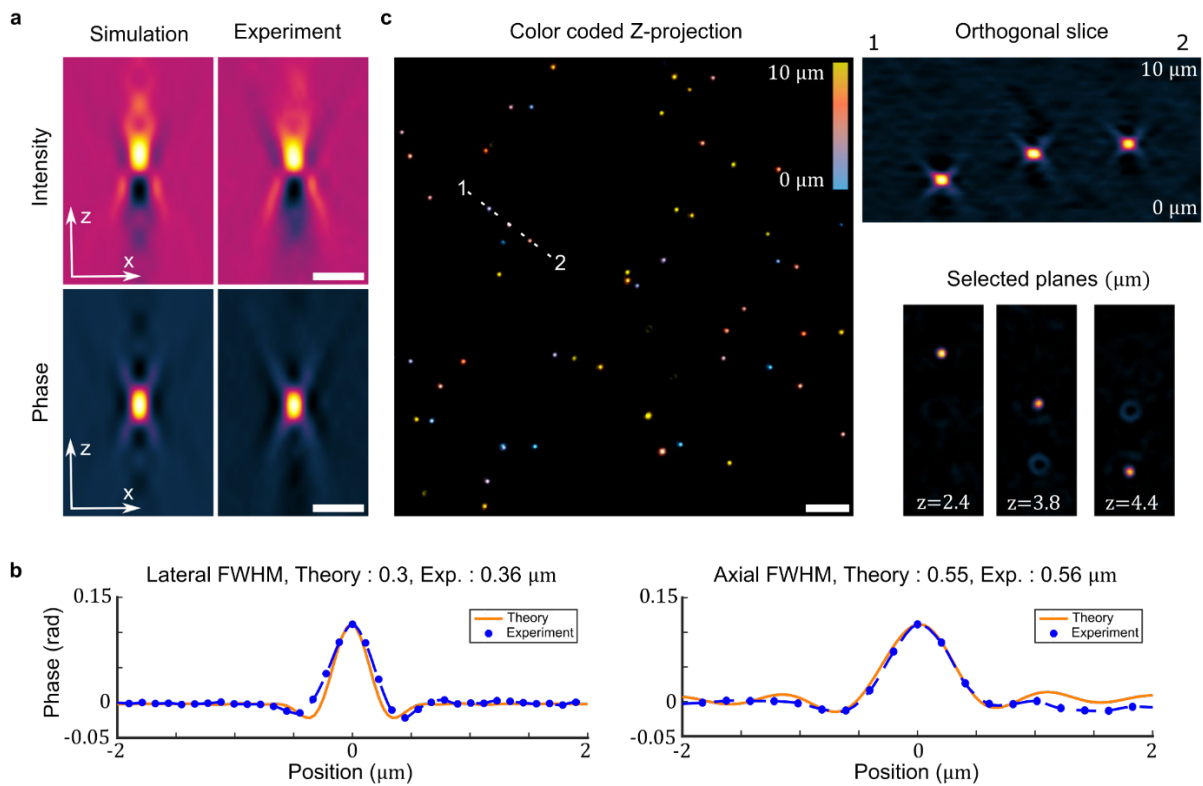

Figure S3 Experiments and simulations with polystyrene beads (a) 3D axial slice of computed and experimental (average over 15 measurements) intensity and phase of 200 nm polystyrene beads. Scalabar 1 μm, (b) Comparison of the theoretical and experimental phase profile showing almost perfect agreement, (c) Color coded maximum z-projection of the phase of sparsely distributed beads. The orthogonal slice AB demonstrates the sectioning capability of the method. Scalabar 5 μm.

## 4 Quantitative phase imaging of a technical sample

Our quantitative phase algorithm is also appropriate for investigating the topology of technical samples. As an example, we took a glass (borosilicate) wafer chemically etched multiple times to create a staircase-like nanometric structure (see Section 8.5).

We imaged the sample with our PRISM microscope (see Section 6), acquiring a total of 8 planes with an inter-plane distance of 350nm and an exposure time of 20ms. The results of this quantitative imaging are shown in Fig. S4.

Fig. S4a shows the brightfield intensity as recorded at plane #4. The in-focus intensity shows a limited contrast. We processed the intensity stack with our algorithm as described in 2.1 to recover a high-contrast quantitative phase image. We show in Fig. S4b the quantitative phase corresponding to the same plane #4.

The staircase structure with various step heights measured in transmission provides a signal caused by all edges. In our transmission configuration, homogeneous structures perpendicular to the optical axis do not scatter. Near each step, we have an air-glass ($n_{air} = 1$, $n_{glass} = 1.51$) interface where the scattering potential is negative in the air $\left( n_{air}^2 - \overline{n^2} < 0 \right)$ and positive in the glass $\left( n_{glass}^2 - \overline{n^2} > 0 \right)$.

# Combined Multi-Plane Tomographic Phase Retrieval and Super-Resolution Optical Fluctuation Imaging for 4D Cell Microscopy - Supplementary Material

As a consequence, the quantitative phase appears as a high-pass filtered version of the underlying object. Therefore, the sign of the differential signal encodes the up-hill/down-hill structure and the total amplitude of the phase encodes the step height. We show in Fig. S4c a line plot comparison between the raw quantitative phase (Fig. 4b, orange line), Atomic Force Microscopy (AFM Fig.4d, blue line) and the "unwrapped" phase (Fig. 4e, dashed green line). The unwrapping is performed by applying an inverse-Laplacian filter to the raw quantitative phase (Fig. 4b) and is used here as a cosmetic tool to help visualize this specific technical sample.

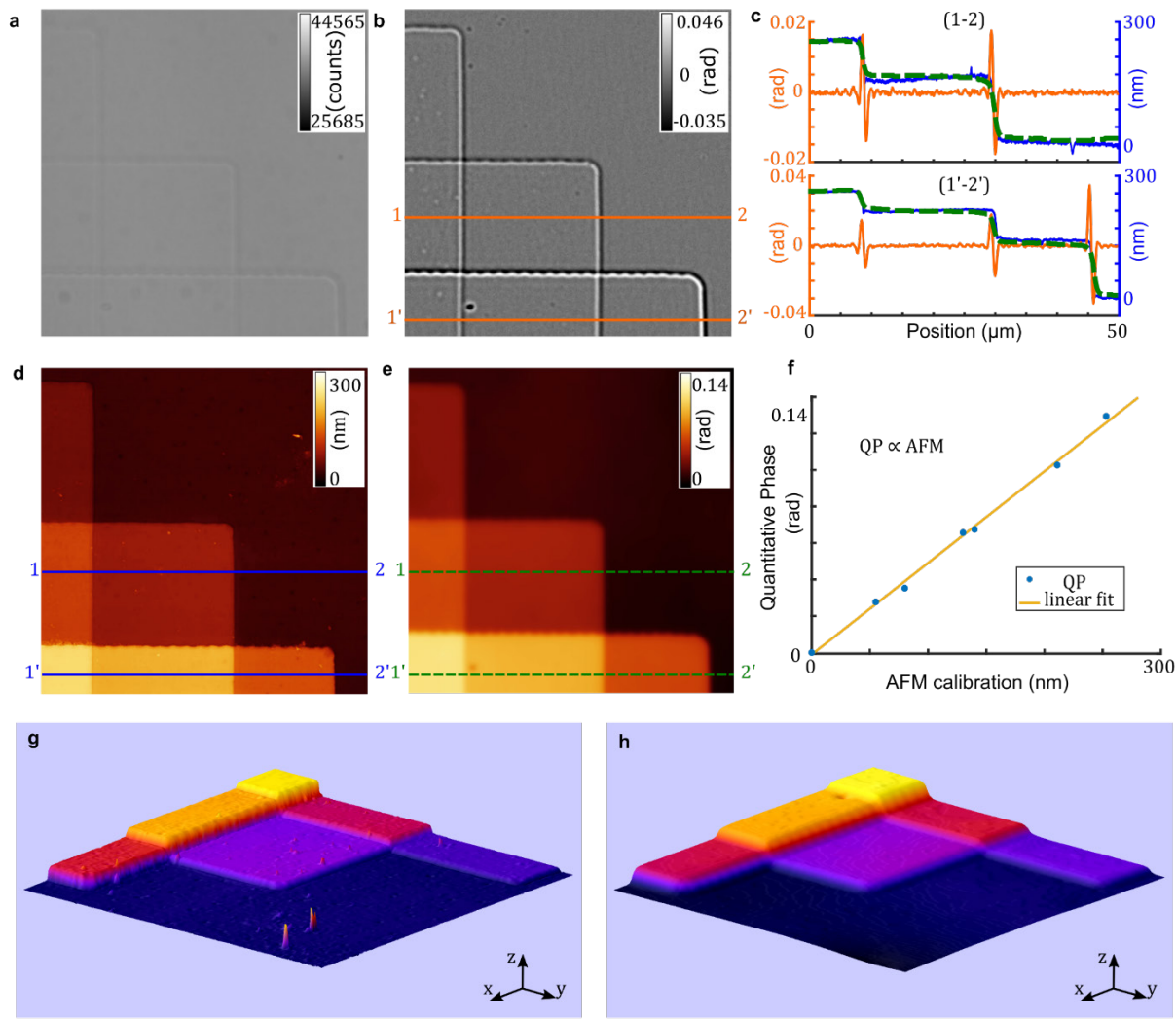

Figure S4 Quantitative phase imaging of technical sample. (a) Brightfield image acquired using the multi-plane configuration (plane #4), (b) Corresponding quantitative phase after processing of the intensity stack, (c) Line plots of selected lines indicated in (b, d and e, (orange) quantitative phase, (blue) AFM measurement, (dashed green) unwrapped quantitative phase, (d) AFM measurement, (e) Inverse-Laplacian filtered quantitative phase, (f) Quantitative phase vs. AFM showing a linear relationship $\left( QP\ [rad] = 0.0005 \left[\frac{1}{nm}\right] * AFM[nm] \right)$, (g, h) 3D surface rendering of the AFM/quantitative phase measurement (using ImageJ[10]).

Fig. S4d shows the AFM measurement; Fig. S4e shows the "unwrapped" quantitative phase obtained by processing Fig. S4b with an inverse-Laplacian filter. The homogeneous refractive index (known a priori)

**Combined Multi-Plane Tomographic Phase Retrieval and Super-Resolution Optical Fluctuation Imaging for 4D Cell Microscopy - Supplementary Material**

was used for calibrating the measured phase against the known step heights. The result is shown in Fig. S4f, which confirms the perfect linear relation when comparing our signal with an independent AFM measurement. In addition, we show a 3D surface rendering of the AFM and quantitative phase measurements (see Fig. S4g, h).

This technical sample analysis underlines the quantitative nature of our method as well as the ability to perform quantitative phase imaging using our multi-plane microscope.

## 5 Fourier space sampling

From a fundamental point-of-view, the quality of our phase retrieval method depends on two main parameters: the inter-plane distance and the total number of planes used for imaging. The inter-plane distance $\Delta z$ is directly related to the Nyquist-Shannon sampling theorem[11]. The number of planes determines how well one can filter the overlap between $\Gamma_+$ and $\Gamma_-$.

In detail, for a given number of plane, the Fourier sampling step is given by $\Delta g_z = \frac{\pi}{(N_{planes}-1)\Delta z}$.

Reducing the number of plane results in a coarser sampling of the Fourier space. This means that the cutoff filter $g_{z,c}$ (equation (S24)) cannot always be optimally applied. Therefore, some useful signal is removed, ultimately resulting in a (further) lateral high-pass filtering of the quantitative phase.

We illustrate in Fig. S5a,b the impact on the phase retrieval when taking $N_{planes} = 3$, 8 or 18 z-planes (scanning inter-plane distance of $\Delta z = 350\ nm$).

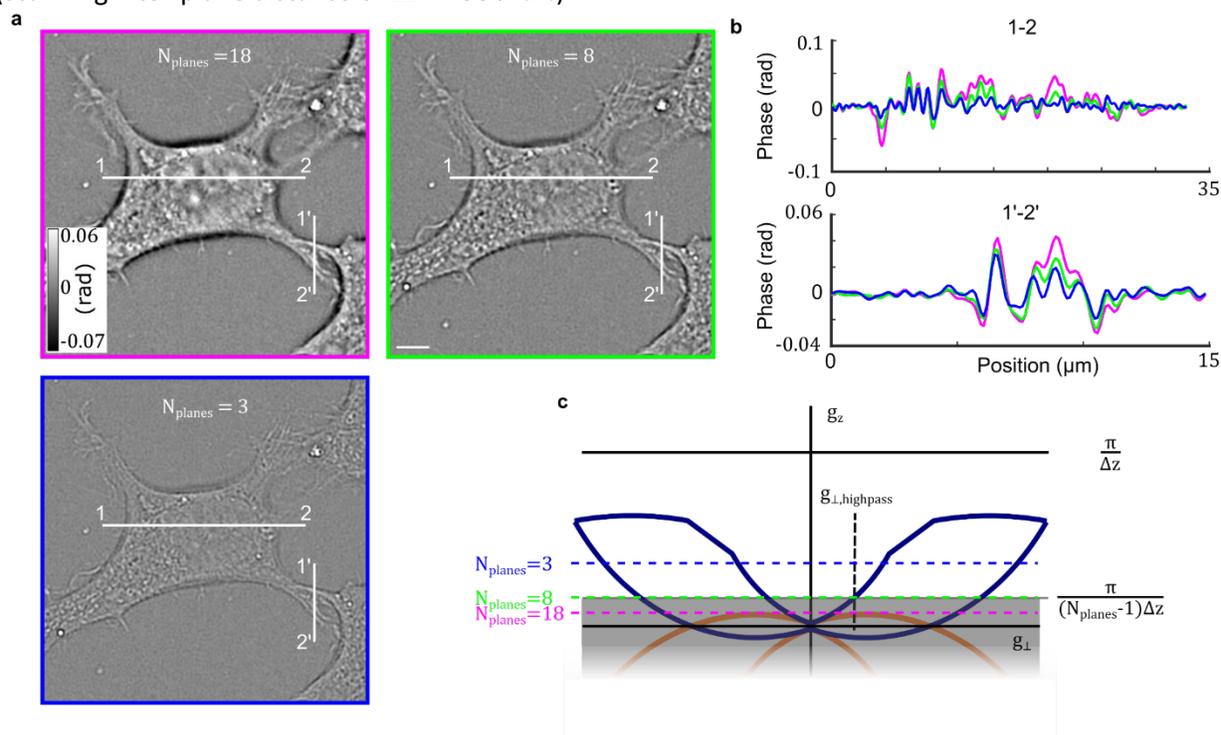

Figure S5: Influence of the number of planes on phase reconstruction (a) $N_{planes}$ = 18,8,3 plane reconstructions of a fixed HeLa cell imaged with an inter-plane distance of $\Delta z = 350$ (magenta, green and blue), highlighting the high-pass filtering imposed by the sampling. By increasing the number of planes, larger structures are clearly revealed (b), Line plots along perpendicular directions indicated in (a) for phase reconstructions with three different samplings showing the consistency of the recovered phase, (c) Theoretical representation of the transfer function $H(k)$ showing how the inter-plane distance and the number of

**Combined Multi-Plane Tomographic Phase Retrieval and Super-Resolution Optical Fluctuation Imaging for 4D Cell Microscopy - Supplementary Material**

planes limits the accessible frequency space. The indicated $N_{planes}$=3,8,18 dotted-lines illustrate the lower accessible $g_z$ limit. Scalebar 5 µm.

The minimal number of planes required for optimal removal of the overlap contribution can be easily estimated from a Fourier sampling analysis and correspond to the case where the sampling step $\Delta g_z$ is equal to the cutoff frequency $g_{z,c}$. A simple estimation supposing a $\Delta z \approx 350\ nm$ and the cutoff limit for suppressing the overlap ($NA_{ill} \approx 0.26$, $k_{0,max} = 11.5 [\mu m]^{-1}$, equation (S24)) results in

$$N_{planes} = 1 + \frac{\pi}{\Delta z\ g_{z,c}} \approx 18\ planes \tag{S28}$$

The high-pass filtering side-effect of limited axial Fourier sampling can be easily seen in Fig. S5c (intersection of the horizontal line $\Delta g_z$ with the transfer function footprint), and the lateral cutoff frequency can be expressed as

$$g_{\perp,highpass} = NA_{ill} k_{0,min} - \sqrt{g_{z,c} - k_{0,min}\sqrt{1 - NA_{ill}^2}}. \tag{S29}$$

We provide the reader equation (S28) as a simple tool to estimate the number of planes required to achieve high-quality phase reconstruction for a given sampling step $\Delta z$, illumination NA and source spectrum (which both influence the term $g_{z,c}$).

In our case, we propose the use of a multi-plane microscope (see Section 6), providing a total of 8 defocused planes. We state the 8 planes as a reasonable compromise between the speed, the low-frequencies recovery and the ability to combine our novel quantitative phase imaging with 3D Super-resolution Optical Fluctuation Imaging (SOFI).

## 6    PRISM multi-plane platform for 3D phase and SOFI imaging

The telecentric multi-plane platform (MP) allows the simultaneous acquisition of 8 fluorescence or brightfield images originating from 8 object planes with a constant inter-plane distance (Fig. S6). The fluorescence excitation is realized by wide-field epi-illumination, whereas phase imaging uses the Koehler brightfield arm from a commercial Zeiss microscope. The detection system is common to both imaging modalities, i.e., the setup is a classical microscope with an integrated telescope containing an image splitting prism in the detection path (optical design (Section 6.3) and specification LOB-EPFL, Lausanne; manufacturing Schott SA, Yverdon, Switzerland). An adjustable field stop in the intermediate image plane prevents the overlap of the images on the cameras. All lenses and optomechanics are standard commercial components, except for the custom-made holder for the prism and cameras (see Fig. S7 and Table S1 for the list of optical components). This versatile MP microscope allows diffraction limited imaging for all 8 conjugated object-image planes.

# Combined Multi-Plane Tomographic Phase Retrieval and Super-Resolution Optical Fluctuation Imaging for 4D Cell Microscopy - Supplementary Material

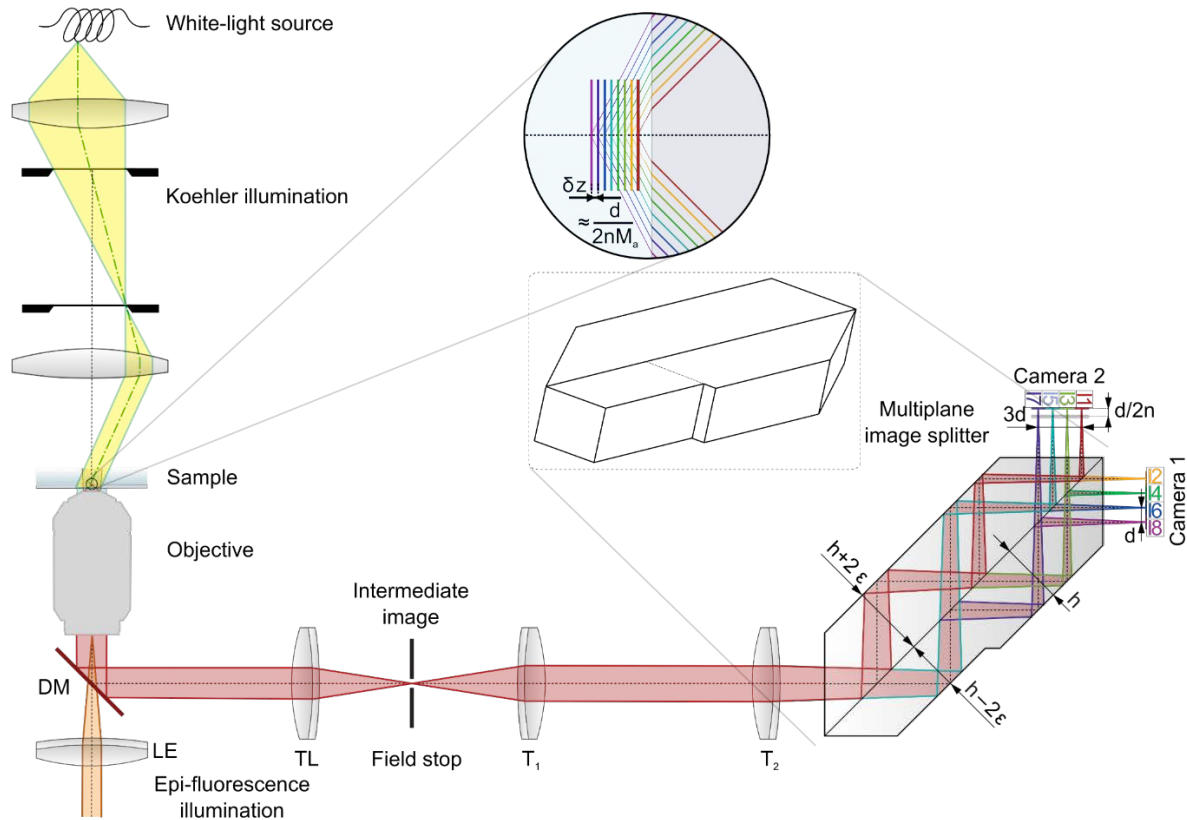

Figure S6. Microscope setup. Wide-field epi-illumination for fluorescence imaging combined with a Koehler trans-illumination for phase imaging. The multi-plane image splitting prism directs the light coming from 8 axially distinct planes in the sample towards 4 adjacent fields of view on each of the two cameras. The lenses in the detection path are arranged as a sequence of four 2f configurations. DM dichroic mirror, LE lens, TL tube lens, $T_1$ and $T_2$ lenses.

| Component | Specifications |
|---|---|
| White light Koehler illumination | Axiovert 100 M, 12 V 100 W Halogen lamp (Carl Zeiss) |
| Fluorescence widefield epi-illumination | 120 mW, 405 nm laser (iBeam smart, Toptica) |
| | 800 mW, 635 nm laser (MLL-III-635, Roithner Lasertechnik) |
| | 800 mW, 532 nm laser (MLL-FN-532, Roithner Lasertechnik) |
| Objective | UPLSAPO 60XW 1.2 NA (Olympus) |
| Microscope stage | piezoLEGS® stage (3-PT-60-F2,5/5) and Motion-Commander-Piezo controller (Nanos Instruments GmbH) |
| Filters | Dichroic zt405/488/532/640rpc, (Chroma) |
| | Emission filters adapted to the respective experiment |
| Tube lens | $f_{TL} = 140\ mm$ (achromatic doublet G036-146-000 (AC 140/22.4, ANR 574070), Qioptiq) |
| Telescope lenses | $f_{T_1} = 160\ mm$ and $f_{T_2} = 200\ mm$ (achromatic doublets G630-631-660 (AC 160/22.4, ANR 576447) and G063-148-000 (AC 200/22.4, ANR 556031), Qioptiq) |
| Prism | Corning C-7980, n= 1.458 (at 587.56 nm), Abbe number V = 67.8 |
| | See Section 6.2 |
| Camera | Orca Flash 4.0 (Hamamatsu), pixel pitch $P_{Camera} = 6.5\mu m$ |

Table S1. Components of the MP detection. All optomechanics are standard commercial components, except for the custom-made holder for the prisms and cameras.

# Combined Multi-Plane Tomographic Phase Retrieval and Super-Resolution Optical Fluctuation Imaging for 4D Cell Microscopy - Supplementary Material

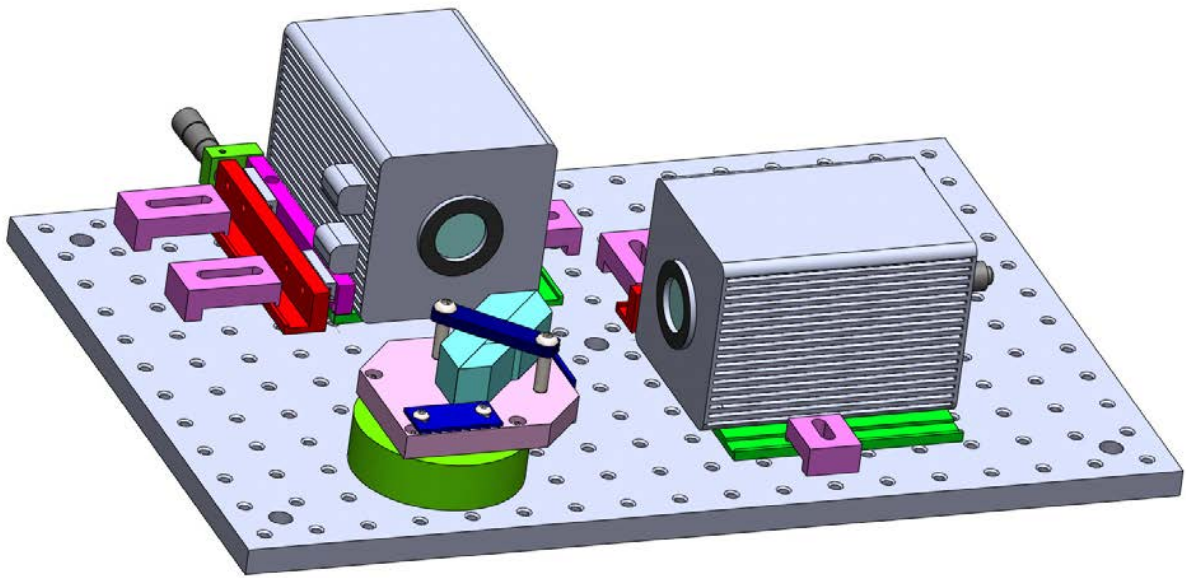

Figure S7. Custom-made holder for the prism and cameras.

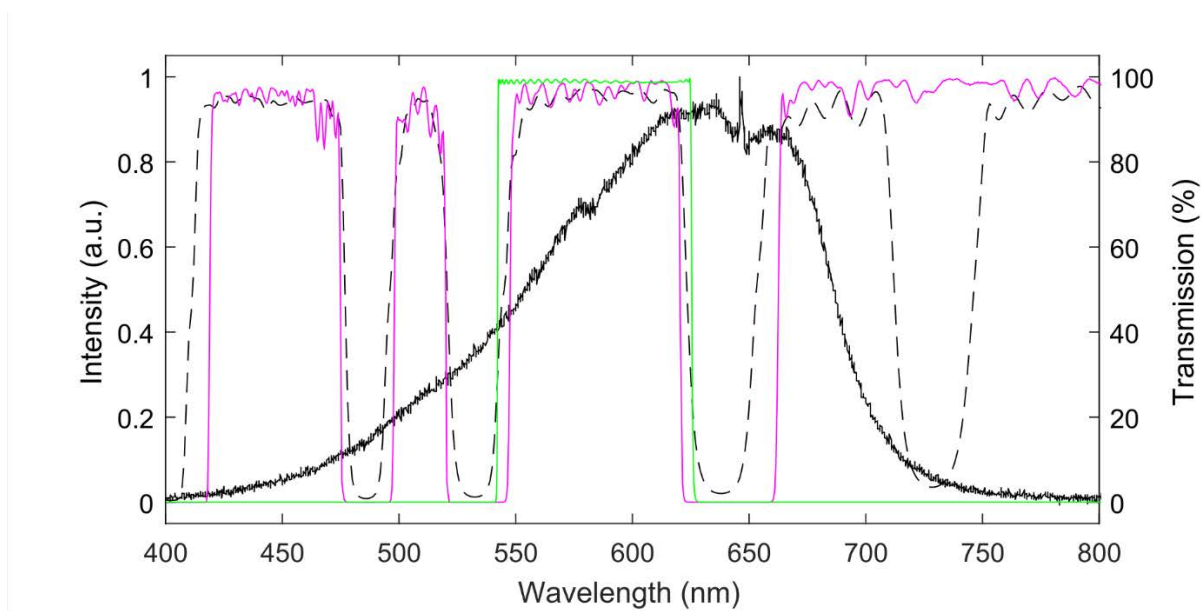

Figure S8. Spectrum of the white-light source and filter set. The spectrum of the halogen lamp (black line) was measured at 4 V drive voltage with a fibre-coupled spectrometer (CCS200M, Thorlabs). The transmission window(s) of the filters are indicated. Dichroic ZT 405/488/532/640/730rpc (dashed black, Chroma), emission filter Bright Line 582/75 (green, Semrock) and quad band emission filter ZET 405/488/532/640m (magenta, Chroma).

**Combined Multi-Plane Tomographic Phase Retrieval and Super-Resolution Optical Fluctuation Imaging for 4D Cell Microscopy - Supplementary Material**

## 6.1 ABCD description of the MP detection

Fig. S9 shows the overall MP concept along with a basic ABCD matrix description for the system. All optical distances are chosen at the corresponding focal length per element such to ensure telecentricity in the object and image spaces. The intermediate image contains the field stop, which prevents overlap between the images at the detector plane(s). The multi-plane image splitter is located in the convergent path and introduces both a lateral offset and an axial path length difference among the image fields. Each object plane has its corresponding conjugated image plane. The overall lateral magnification of the platform is that of the classical microscope combined with a telescope $M_l = M_{Mic} M_{tel} = \frac{f_{TL} f_{T_2}}{f_{Objective} f_{T_1}}$.

For this study, we chose a 60× water-immersion objective and lenses as indicated in Table S1. Altogether, this amounts to a lateral magnification of $M_l \approx 58.3$ which results in a back-projected pixel size $p = \frac{p_{Camera}}{M_l} \approx 111 \, nm$.

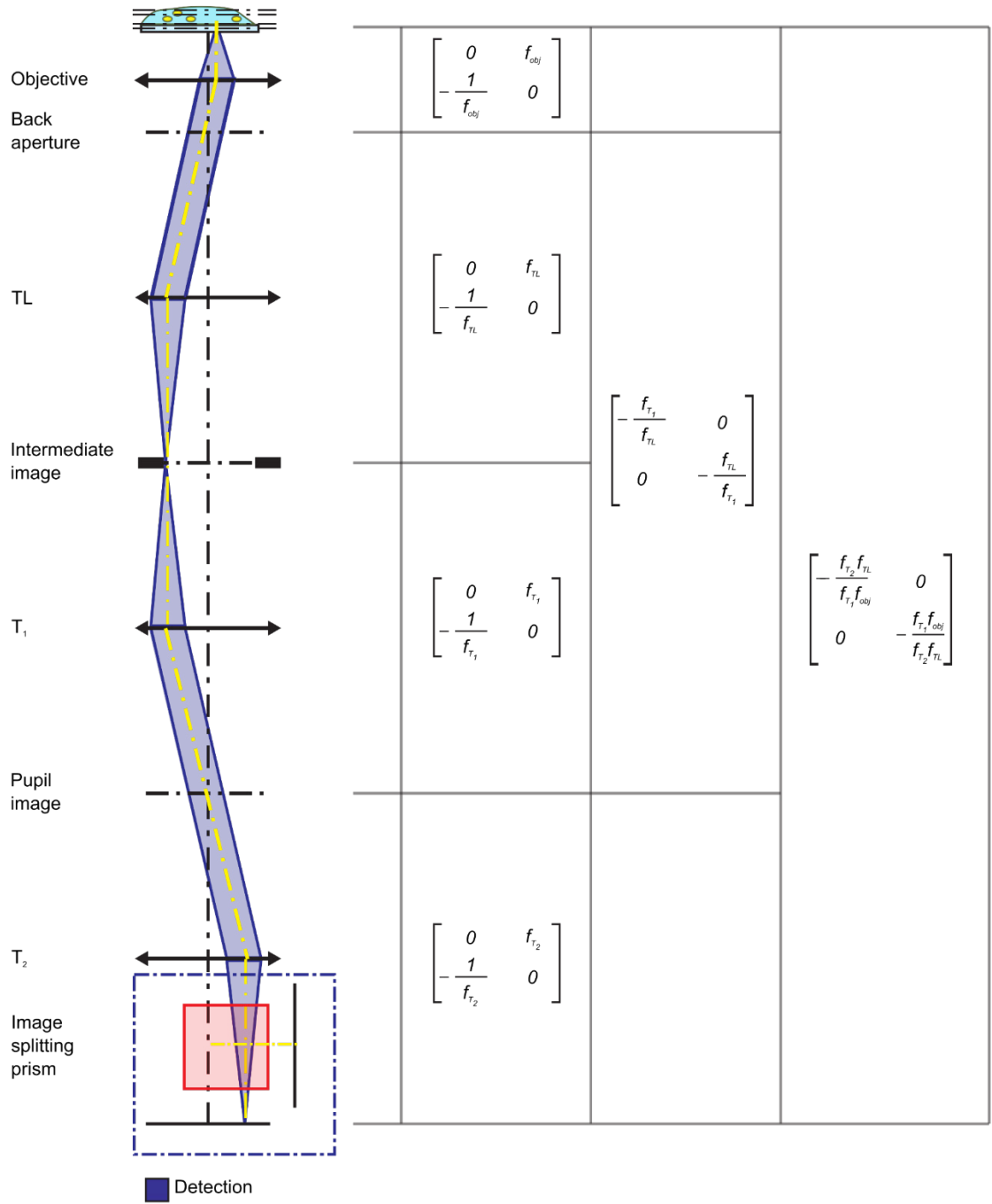

Figure S9. ABCD description of the MP concept. The full system can be decomposed into 4 distinct 2f systems. The scheme shows the system as an object- and image-side telecentric system. The TL-T$_1$ system is a telescope, mapping the backfocal plane of the objective into the pupil plane. We use the convention $\begin{bmatrix} h \\ \alpha \end{bmatrix}$ with the ray height $h$ and ray angle $\alpha$.

# Combined Multi-Plane Tomographic Phase Retrieval and Super-Resolution Optical Fluctuation Imaging for 4D Cell Microscopy - Supplementary Material

## 6.2 Image splitter prism design

As shown in Fig. S6 and S10, the image splitter consists of 3 individual prisms glued along the common interfaces. The different images have individual paths and accumulate path length differences such that a conjugated object-image condition is ensured. The light paths undergo multiple total reflection at the outer prism interfaces, whereas the common inner interface has a customized 50:50 coating for equal image intensities.

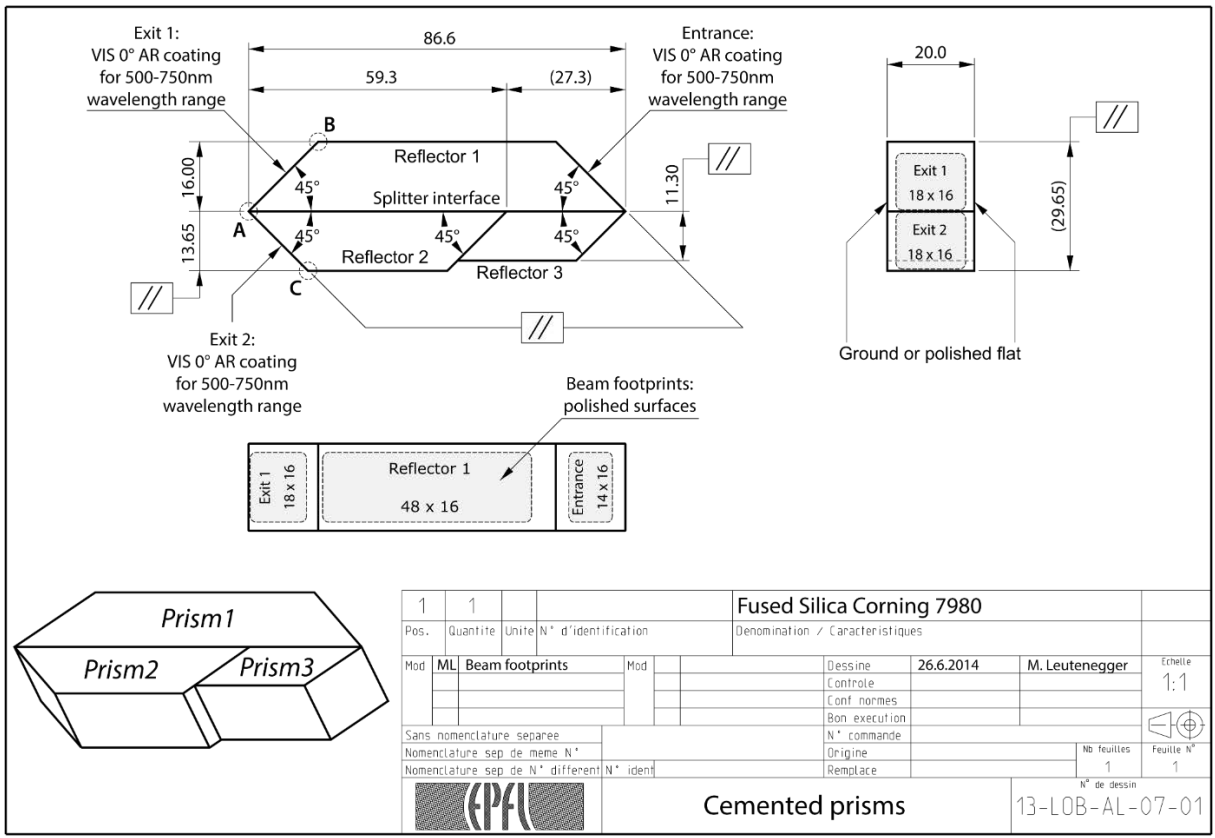

Figure S10. Technical drawing of the image splitter prism assembly.

Essential for a constant inter-plane distance and a constant lateral displacement $d$ are the different prism heights $h$ and $h \pm 2\varepsilon$ ($\varepsilon = \frac{d}{\sqrt{2}}$). As indicated in Fig. S6, the images are channeled in an interlaced fashion, i.e. the odd images (1, 3, 5, 7) are acquired by camera 1, the even images (2, 4, 6, 8) are acquired by camera 2. The neighboring images on both cameras are laterally displaced by a distance $d$, which corresponds to a path length difference of $\Delta z = \frac{d}{n}$ (prism refractive index $n = 1.458$), i.e. for example the path length difference of image 7 to image 5 equals $\Delta z$. The first camera is additionally shifted by $\Delta z_{IP} = \frac{d}{2n}$, which results in a constant inter-plane (IP) distance of the consecutive axial planes $\Delta z_{IP}$. We chose the displacement of neighboring images $d = 3.32$ mm such that the row of four images matches the width of the sCMOS sensor (Orca Flash 4.0, Hamamatsu). The geometric path lengths of the rays in the prism range from 96.52 mm for the images 7 and 8 to 106.48 mm for the images 1 and 2. We performed an analysis of image aberrations using Zemax. Results are summarized in Section 6.3 below.

# Combined Multi-Plane Tomographic Phase Retrieval and Super-Resolution Optical Fluctuation Imaging for 4D Cell Microscopy - Supplementary Material

The inter-plane distance of consecutive axial planes in object space is then given by $\delta z_{IP} = \frac{\Delta z_{IP}}{M_a}$ where $M_a = M_l^2$ is the axial magnification. Experimentally, the inter-plane distance was estimated as $347 \pm 11$ nm (mean $\pm$ standard deviation of 8 individual measurements of surface immobilized fluorescent beads scanned along the optical axis in 200 nm steps).

## 6.3 Analysis of the MP design performance

The overall system has been fully designed based on ray tracing with Zemax (Radiant Zemax LLC). The layout of the optical design of the detection path is shown in Fig. S6. As we are lacking the objective's design data, we modelled the objective as a paraxial lens with 3 mm focal length and 7.2 mm aperture diameter. The field stop is adjusted to suppress crosstalk between the different images, resulting in an image side numerical aperture of $NA_i = 0.031$. All lenses (TL, $T_1$, $T_2$) are in this paraxial path and are standard lenses (Qioptiq) as indicated in Table S1. All optical distances are chosen at the effective focal length per element (see Fig. S11). The objective is placed in a telecentric configuration and the tube lens TL to the image plane is almost image-side telecentric (exit pupil position > 33 m from the image plane).

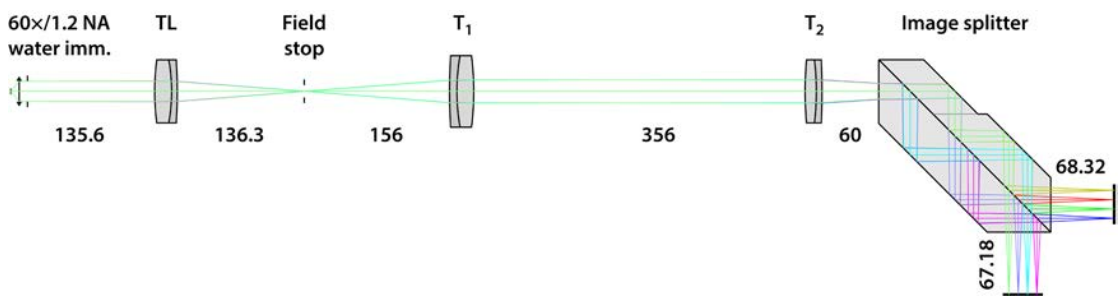

Figure S11. Layout of the detection system containing a 60×/1.20 NA water immersion objective, the tube lens ($f_{TL} = 140$ mm), the field stop, the telescope lenses $T_1$ ($f_1 = 160$ mm) and $T_2$ ($f_2 = 200$ mm) and the image splitter with two cameras. The free-space distances are indicated in millimeters and are 3x downscaled in the Figure.

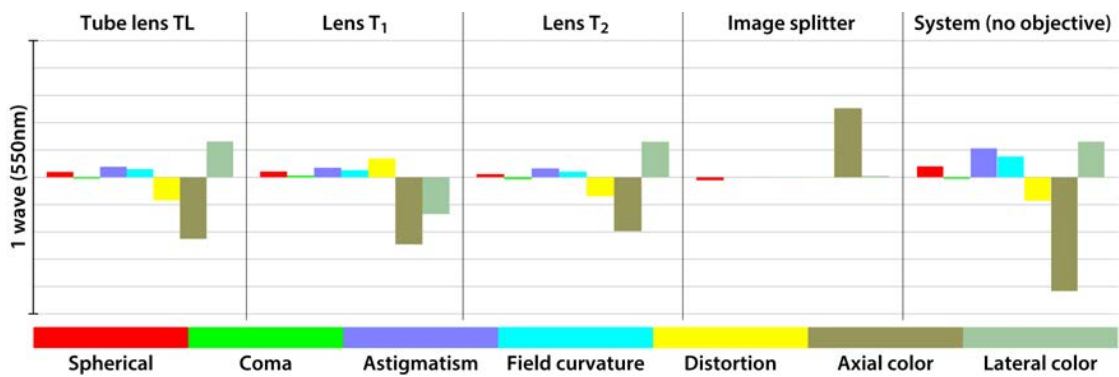

Figure S12. Seidel diagram (third-order aberrations) of the detection path. The major aberrations of the system are axial color with about 0.4 waves, and lateral color and astigmatism with about 0.1 waves each.

Our analysis of the optical aberrations based on the Seidel diagram (Fig. S12) shows that the spherical aberration due to the prism is insignificant. Moreover, the axial color (chromatic length aberration, CLA) of the prism compensates the axial color of all lenses. Therefore, the remaining system aberrations are due to the "unknown" aberrations of the objective and the residual aberrations of the lenses. The Seidel diagram shows that the axial color is the dominant aberration with CLA < λ/2 in the wavelength range

## Combined Multi-Plane Tomographic Phase Retrieval and Super-Resolution Optical Fluctuation Imaging for 4D Cell Microscopy - Supplementary Material

500–650 nm. These residual aberrations are even lower for the used wavelength range. The optical path length differences (OPD) are below λ/2 for the wavelength range and field sizes up to 100 μm in diameter (Fig. S13). For all fields and wavelengths, the geometrical PSF is below the Airy radius indicating clearly a diffraction-limited performance (Fig. S14). As shown in the spot diagrams, the chief ray positions shift slightly due to lateral chromatic aberration well below the Airy radius, indicating that these residual lateral chromatic aberrations are insignificant. The diffraction-limited performance of the detection system is also evidenced by the polychromatic modulation transfer and point spread functions (Fig. S15), showing a Strehl ratio > 0.95.

In summary we have a diffraction limited performance of our MP system.

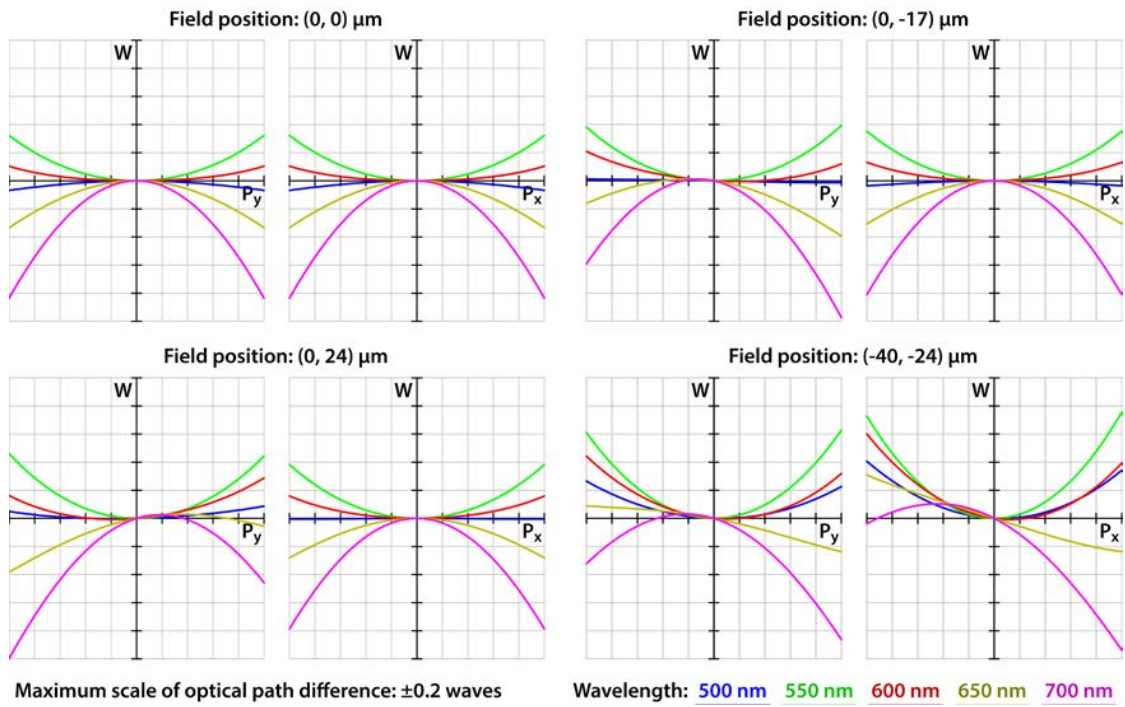

Figure S13. Optical path length difference (OPD) versus normalized pupil position. The OPD shows diffraction-limited performance ($|W| < 0.2$ waves) over an extended wavelength range and field.

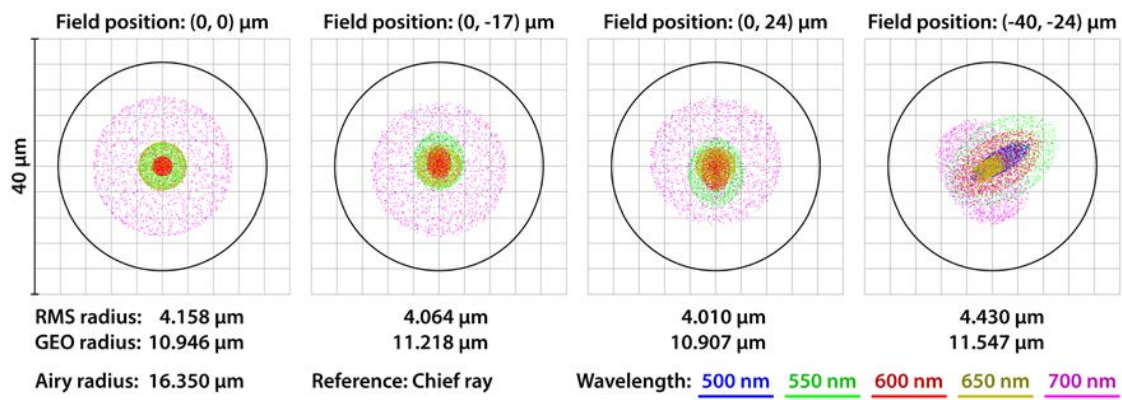

Figure S14. Spot diagrams of the ray intersections in all image planes. The root-mean-square (RMS) radii of the spots are much smaller than the Airy radius, which confirms the diffraction-limited performance across all image fields.

**Combined Multi-Plane Tomographic Phase Retrieval and Super-Resolution Optical Fluctuation Imaging for 4D Cell Microscopy - Supplementary Material**

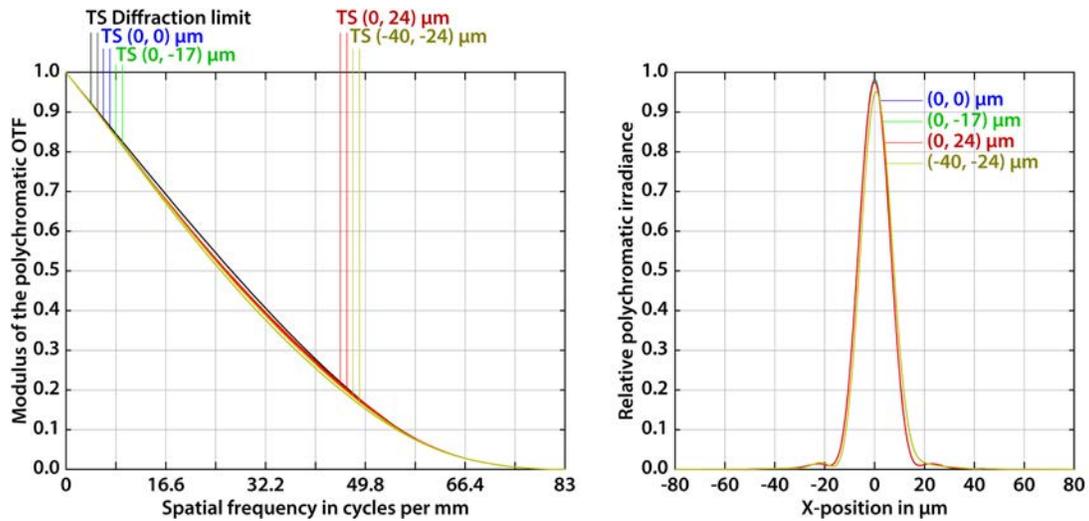

Figure S15. Polychromatic modulation transfer function (MTF) and point spread function (PSF). The MTF closely approaches the diffraction limit for all field positions and the PSF peak value and the Strehl ratio – are well above 0.8.

### 6.4 MP prism calibration

3D multi-plane imaging demands accurate calibration of the image planes. Co-alignment for both imaging modalities is based on an affine transformation and bilinear image interpolation (see Fig. S16). The transformation parameters are extracted from a calibration measurement of fluorescent beads scanned along the optical axis in 200 nm steps over the whole sampling volume.

The beads calibration measurements are also used to correct the transmission variation among the 8 image channels (slight deviation from 50:50 channel splitting) for SOFI processing. For phase imaging, the channels transmission correction is based on brightfield images.

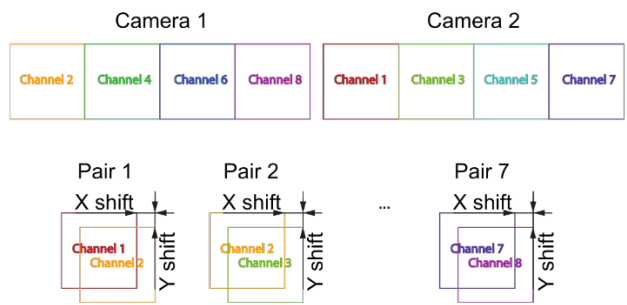

Figure S16. Co-registration of image planes. An affine transformation and bilinear interpolation based on a calibration measurement with fluorescent beads is applied to (pairs of) image channels at different steps in the analysis routine.

**Combined Multi-Plane Tomographic Phase Retrieval and Super-Resolution Optical Fluctuation Imaging for 4D Cell Microscopy - Supplementary Material**

## 7 Workflow - MP SOFI analysis

Fig. S11 shows the MP SOFI analysis workflow step by step.

1. Acquisition of raw images followed by image processing in subsequences. This avoids bleaching induced correlations over the full raw image sequence[12–14].
2. 3D Cumulant analysis at zero-time lag in a sliding bi-plane configuration for minimizing interpolation induced noise; Bi-plane cumulant block co-alignment.
3. Cumulant flattening, deconvolution and linearization followed by subsequence linear combination[15,16].

More details about the algorithm can be found in [12].

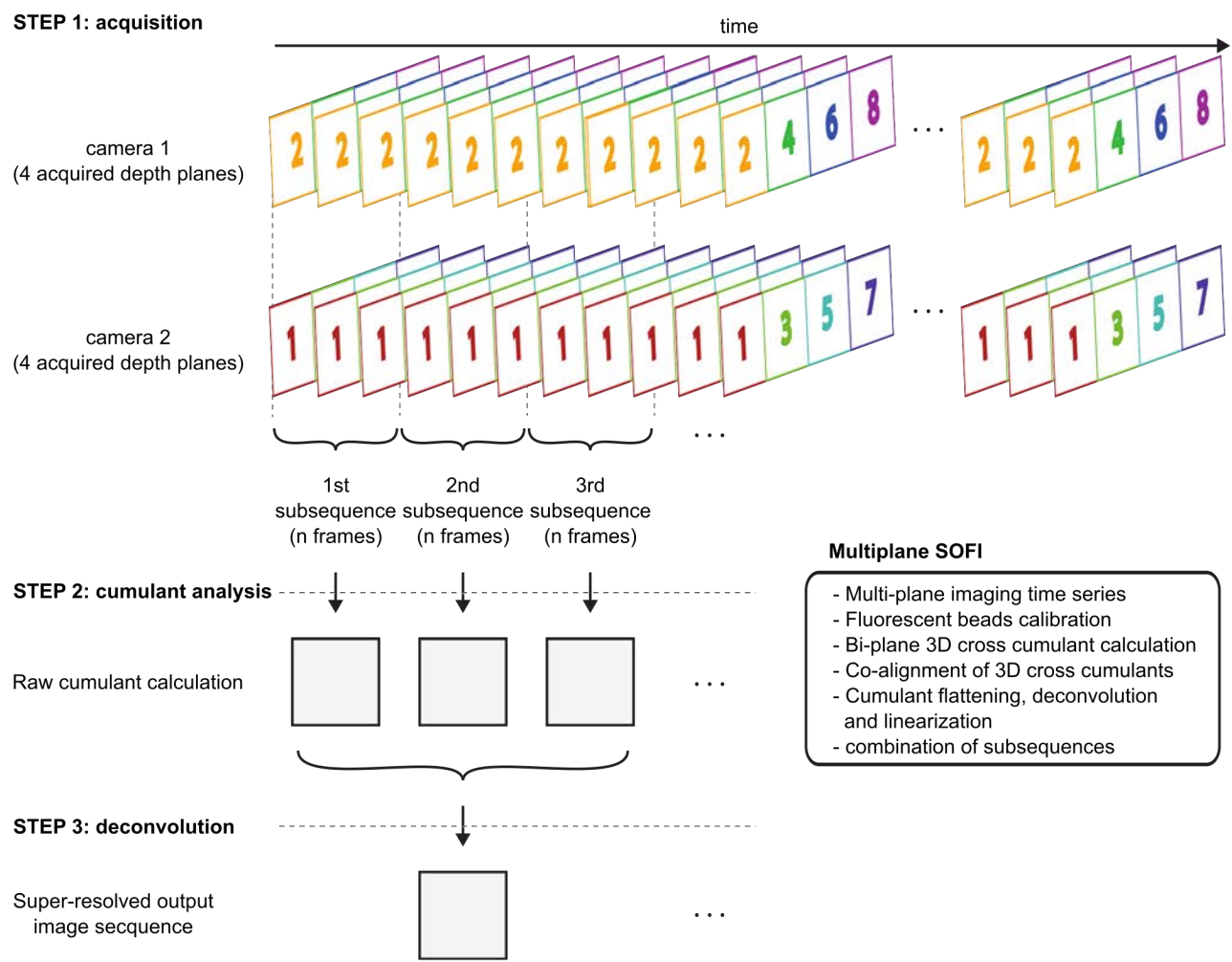

Figure S17. Workflow MP SOFI analysis.

# 8 Experiments

## 8.1 Imaging and Analysis Parameters

| Experiment | Figure | Imaging mode | Excitation intensities[a] | Filter | Total # frames | Inter-plane distance | Frame rate[e] |
|---|---|---|---|---|---|---|---|
| HeLa cell fixed, unlabelled | Fig. 2, Fig. S4 | phase | 4 V halogen lamp | Bright Line 582/75[c] | 50[d] | 200 nm | 20 ms |
| Human fibroblast cells live, unlabelled | Fig. 4 | phase | 4 V halogen lamp | Bright Line 582/75[c] | 5000 | 350 nm | 200 Hz |
| HeLa cell – Alexa Fluor 647 immunolabeled microtubules | Fig. 5 | SOFI | <2 kW cm$^{-2}$ at 635 nm 7 W cm$^{-2}$ at 405 nm | ZET405/488/532/640nm[b] | 5000 | 350 nm | 50 Hz |
| | | phase | 4 V halogen lamp | | 1 | 350 nm | 20 ms |
| primary hippocampal neuron – Alexa Fluor 647 immunolabeled α –synuclein aggregates | Fig. 5 | SOFI | <2 kW cm$^{-2}$ at 635 nm 7 W cm$^{-2}$ at 405 nm | ZET405/488/532/640nm[b] | 5000 | 350 nm | 50 Hz |
| | | phase | 4 V halogen lamp | | 1 | 350 nm | 20 ms |
| RAW 264.7 cells expressing Lifeact-Dreiklang | Fig. 5 | SOFI | 800 W cm$^{-2}$ at 532 nm 8.5 W cm$^{-2}$ at 405 nm | Bright Line 582/75[c] | 3000 | 350 nm | 50 Hz |
| | | phase | 4 V halogen lamp | | 1 | 350 nm | 20 ms |
| primary hippocampal neuron – Alexa Fluor 647 immunolabeled α –synuclein aggregates, | Fig S1 | phase | 4 V halogen lamp | Bright Line 582/75[c] | 200[d] | 200 nm | 20 ms |
| HeLa cell, unlabelled | Fig S2 | phase | 4 V halogen lamp | Bright Line 582/75[c] | 1 | 350 nm | 20 ms |
| Polystyrene beads, 200nm | Fig. S3 | phase | 4 V halogen lamp | Bright Line 582/75[c] | 50[d] | 200 nm | 20 ms |
| Human embryonic kidney HEK 293T cells, murine macrophage RAW 264.7 cells, mouse hippocampal primary neuron, human fibroblast, all unlabeled except for Alexa Fluor 647 immunolabeled α –synuclein aggregates in primary hippocampal neuron | Fig. S12 | phase | 4 V halogen lamp | Bright Line 582/75[c] | 30-70[d] | 200 nm | 20 ms |
| HeLa cell, live | Fig. S13 | phase | 4 V halogen lamp | Bright Line 582/75[c] | >60 | 350 nm | 2 Hz |
| HeLa cell expressing Vimentin-Dreiklang | Fig. S14 | SOFI | 800 W cm$^{-2}$ at 532 nm 8.5 W cm$^{-2}$ at 405 nm | Bright Line 582/75[c] | 1500 | 350 nm | 50 Hz |
| | | phase | 4 V halogen lamp | | 5000 | 350 nm | 50 Hz |

[a]measured near the back focal plane of the objective for fluorescence illumination, [b]Chroma, [c]Semrock, [d]z-stack stage scanning, all other experiments were performed with multi-plane acquisition, [e]for single frame imaging, the exposure time is indicated

Table S2. Imaging and Analysis Parameters.

## 8.2   Phase imaging

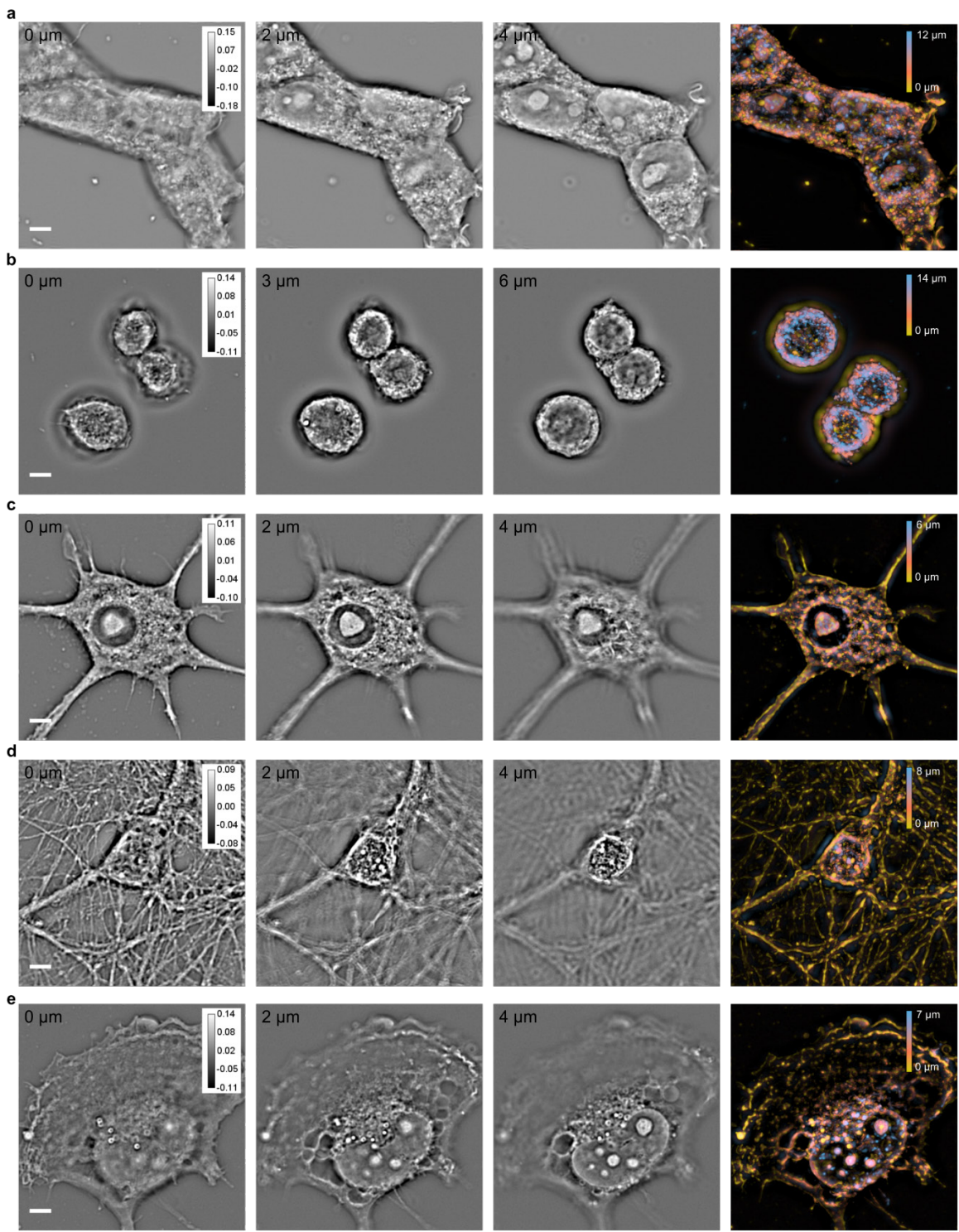

Figure S18: Phase imaging of fixed cell samples using large image stacks. Samples were scanned with $\Delta z = 200$ nm and x-y-slices at different height along with a color-coded maximum intensity-z-projection of images with a phase threshold = 0 are shown for each cell type. Phase in radian. (a) Human embryonic kidney HEK 293T cells, (b) murine macrophage RAW 264.7, (c) LPS stimulated murine macrophage RAW 264.7, (d) mouse hippocampal primary neuron, (e) human fibroblast. Scalebar 5 µm

# Combined Multi-Plane Tomographic Phase Retrieval and Super-Resolution Optical Fluctuation Imaging for 4D Cell Microscopy - Supplementary Material

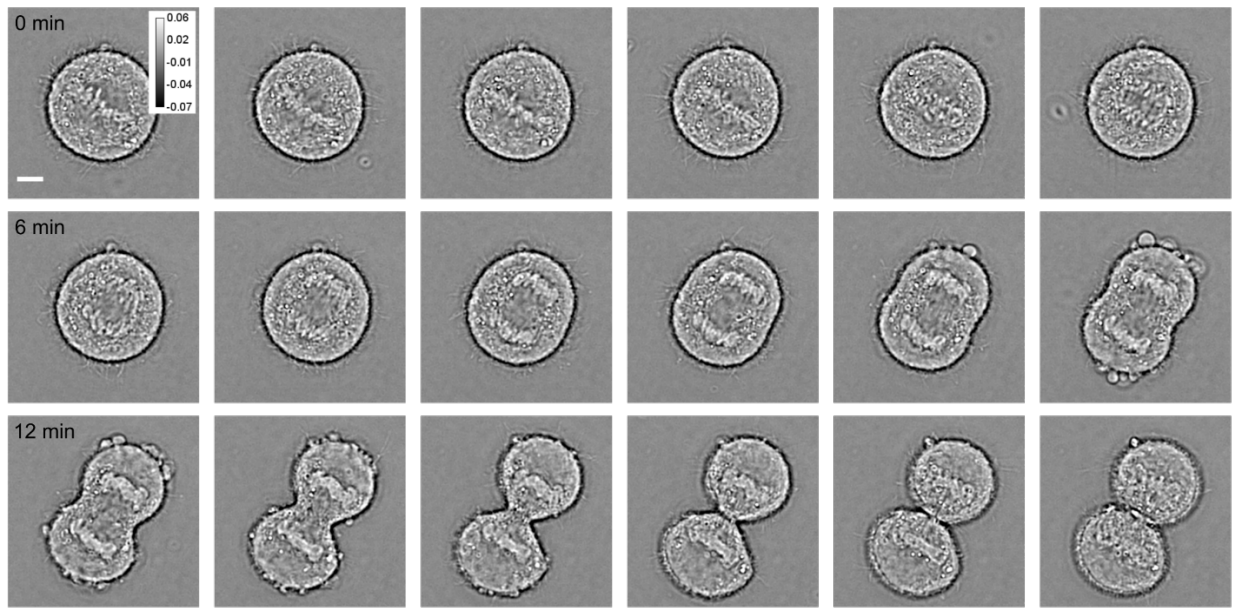

Figure S19: Dividing HeLa cell. Living HeLa cell imaged every 30 s as it undergoes mitosis from metaphase to telophase using the MP phase microscope (every second image is shown for plane 7). The phase is given in radian. Scalebar 5 µm.

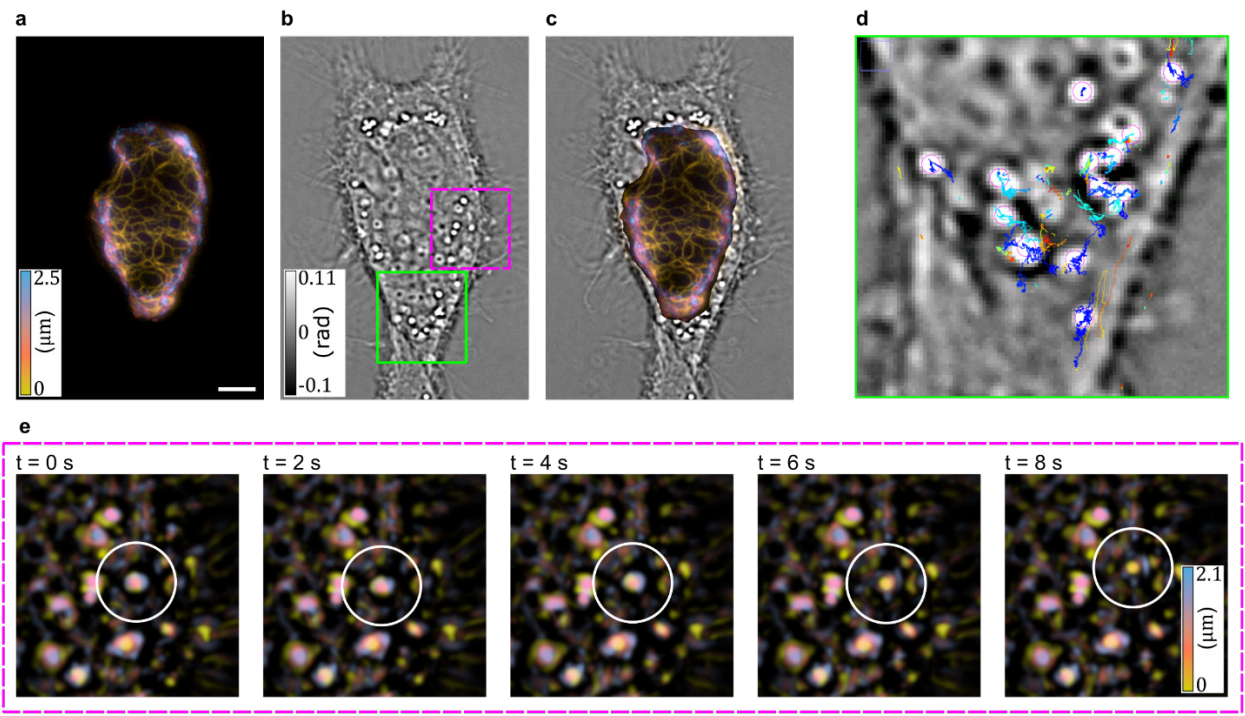

Figure S20 Multi-plane SOFI and dynamic phase imaging. (a) SOFI maximum intensity projection of a living HeLa cell expressing Vimentin-Dreiklang. (b) First frame of a 100 s 3D phase movie imaged at 50 Hz. (c) Overlay of 3D SOFI and phase. (d) Example tracking of highly scattering vesicles in a region of interest (green) using TrackMate. (e) Phase maximum intensity projection (Threshold T = 0 rad) for selected time point displaying 3D dynamics. Scalebar 5 µm.

**Combined Multi-Plane Tomographic Phase Retrieval and Super-Resolution Optical Fluctuation Imaging for 4D Cell Microscopy - Supplementary Material**

### 8.3   Fluorescent beads calibration sample preparation

Fluorescent beads (PS-Speck™ Microscope Point Source Kit, orange and deep red, Thermo Fisher Scientific) with a diameter of 0.175 μm were allowed to dry in a Lab-tek® II chambered cover slide (nunc) and subsequently covered with the provided immersion medium.

### 8.4   Polystyrene beads sample preparation

Polybead® Polystyrene 0.2 Micron Microspheres (2.57 % Solids-Latex) (Polysciences Inc.) were diluted 1:100 in Millipore water followed by 1:100 dilution in 0.5 % agarose solution. 100-200 μl of this mixture were dispersed on a coverslip (0.17 mm thickness (Assistent)) for imaging. An intensity stack of 50 μm x 50 μm x 10 μm (50 planes spaced by 200 nm) has been acquired and processed to recover the corresponding phase.

### 8.5   Atomic Force Microscopy (AFM) sample preparation and measurement

Cleanroom fabrication was performed in the Center of MicroNanoTechnology (CMi, EPFL) following standard protocols:

1. Pyrex/ glass wafers with a thickness of 145 ± 15 μm were coated with aluminum on the backside with a thickness of 500nm.
2. Photolithography: The AZ1512 photoresist was spin-coated on the top of the wafer to a final thickness of 1 μm. The wafer was then baked at 112 °C for 1 min 30 s. The exposure was performed using the MJB4 tool (Süss) for 0.8 s using a custom made mask, followed by development using the developer AZ for 55-65 s.
3. Dry etching: Reactive ion etching was performed using a $C_4F_8$ and $O_2$ based plasma on the LPX machine (SPTS Technologies). The etch time was 30 s and the photoresist was removed by $O_2$ plasma in the same machine by a consecutive processing step for 1 min.

Process 2. and 3. were repeated overall 3 times until the desired pattern was made.

4. The backside aluminum layer was removed by wet etching in ANP solution for 4 min.

AFM imaging was performed with the Dimension Icon (Bruker) in tapping mode in air. The used AFM tip was a RTESPA-525 and the height sensor data image was acquired with 0.5Hz with a resolution of 1024 pixels x1024 pixels on a scan area of 70 μm x 70 μm.